\documentclass[twocolumn,preprintnumbers,amsmath,amssymb,pra,showkeys]{revtex4}

\usepackage{amsmath}
\usepackage{mathbbol}
\usepackage{amssymb}
\usepackage{graphicx}
\usepackage{dcolumn}
\usepackage{bm}

\usepackage{yfonts}
\usepackage{float}
\usepackage{hyperref}
\usepackage{mathrsfs}
\usepackage{color}
\usepackage{pgfplots}
\usepackage{bbold}
\usepackage{filecontents}
\usepackage{mathtools}

\begin{document}

\title{Generation of Spin Cat States in an Engineered Dicke Model}

\author{Caspar Groiseau$^{1,2}$}
\email{cgro288@aucklanduni.ac.nz}
\author{Stuart J. Masson$^{3}$}
\author{Scott Parkins$^{1,2}$}
\affiliation{%
$^1$Dodd-Walls Centre for Photonic and Quantum Technologies, New Zealand\\
$^2$Department of Physics, University of Auckland, Private Bag 92019, Auckland 1010, New Zealand\\
$^3$Department of Physics, Columbia University, 538 West 120th Street, New York, New York, 10027-5255, USA
}%


\date{\today}

\begin{abstract}
We study trajectories of collective spin states of an ensemble of spinors. The spinors considered here are either trapped ions in free space or atoms confined in a cavity, both systems of which are engineered through their interactions with light fields to obey an effective Dicke model. In an appropriate limit of the Dicke model, one obtains one-axis twisting dynamics of the collective spin and evolution after a finite time to a spin cat state, or, in the long-time limit, the Dicke state $|S,0\rangle_x$, conditioned upon there being no photon emissions from the system (i.e., no quantum jumps). If there is a jump, however, the system evolves probabilistically into one of a finite number of entangled-state cycles, where the system then undergoes a persistent sequence of jumps between two Dicke state superpositions in a rotated basis. The different cycles can be distinguished by the frequency at which jumps occur.
\end{abstract}

                       
\maketitle
\section{Introduction}
\label{intro}

A so-called ``cat state'' is a quantum superposition of two quasi-classical states, in analogy to Schr\"odinger's original Gedankenexperiment \cite{schrodinger1935gegenwartige}. As well as this fundamental physics aspect, such states are especially interesting for the field of quantum metrology as they allow for quantum-enhanced measurements \cite{RevModPhys.90.035005}. 
Hence, there is considerable interest in the generation of such states, particularly in the context of light fields and of spin angular momentum in atomic ensembles \cite{2019arXiv190500320S,arXiv1806.05495,ourjoumtsev2007generation,friedman2000quantum,facon2016sensitive,sychev2017enlargement,Thekkadath2020engineering,PhysRevA.99.063813}.
Cat states of light involving quasi-classical states of large average photon number are notoriously difficult to produce because of their extreme sensitivity to even very small losses in propagation and in optical elements. 
Spin angular momentum states of atoms, however, offer better possibilities for mesoscopic or macroscopic cat states due to the generally long lifetimes (and coherence times) of the involved spin states, which are typically stable ground hyperfine states. 

Nevertheless, the preparation of large-sized spin cat states is still very challenging, as decoherence still scales with the size of the system (i.e., the magnitude of the total spin), and evolution that produces a cat state in general requires a collective (all-to-all) interaction amongst the atoms. 
Such dynamics is offered by small, tightly-confined (``single-mode'') spinor atomic Bose-Einstein condensates, but the timescales involved are slow as a result of the relatively weak atom-atom interaction strength, and particle losses are difficult to avoid.   

An alternative, and increasingly prevalent means of implementing collective atom-atom interactions is via the platform of cavity quantum electrodynamics (cavity QED), whereby a cavity light field facilitates the exchange of excitation between atoms, i.e., atom-atom interactions are mediated by the cavity field, and are effectively infinite range in nature.
Prominent amongst examples of such cavity-mediated collective atomic interactions is the Dicke model \cite{garraway_dicke_2011,kirton_introduction_2019}.


Here we explore two alternative ways of preparing cat states of two maximal spin projections using recently implemented effective Dicke models: first for trapped ions \cite{PhysRevA.97.042317,PhysRevLett.121.040503} and then for atom ensembles with cavity-mediated Raman transitions \cite{PhysRevLett.119.213601,Zhiqiang:17}.

The paper is organised as follows: we start in  Section~\ref{sec2} by introducing a somewhat general, open Dicke model for $N$ particles. By considering a dispersive limit of this model, we show how one can obtain a particular, effective dynamics -- the so-called 
{\em one axis twisting} Hamiltonian -- that is known to generate cat states \cite{PhysRevA.56.2249,PhysRevB.57.7474}. With damping, the one-axis twisting dynamics is modified and we derive the non-Hermitian time evolution of the spin wave function by changing to a convenient (rotated) basis. We also introduce the fidelity and quantum Fisher information as measures to quantify the success of our scheme in preparing cat states. 
Then, in Sections \ref{ions} and \ref{cavityqed} we take a closer look at the two physical systems -- trapped ions and atoms in a cavity -- in which our scheme could potentially be put into practice. 
For each of these systems we examine the influence of the most relevant decoherence mechanisms on cat-state generation using our scheme. We then consider the requirements that these decoherence mechanisms put on potential experimental values of the key parameters, enabling us to gauge the feasibility of our scheme.

In Section \ref{cavityqed} we additionally explore quantum jump trajectories, heralded by photon emissions from the cavity, which generate ``kitten'' states (cat states of smaller spin amplitude) in a random fashion. We close with our conclusions in Section \ref{concl}.

\section{One-Axis Twisting Dynamics}
\label{sec2}
\subsection{Engineering from a Dicke model}

We consider an ensemble of $N$ identical spinor particles (of total spin $S=Ns$, where $s$ is the spin of each particle) coupled to a, for now unspecified, bosonic mode, which is in equilibrium with a thermal reservoir of mean excitation number $\bar{\mathfrak{n}}$, and subject to decay at a rate $\kappa$. Throughout this paper $S$ is assumed to be an integer.
The master equation for the system density operator $\hat{\rho}$ is $(\hbar=1)$
\begin{equation}
    \dot{\hat\rho}=-i\left[\hat H,\hat\rho\right]+\kappa(\bar{\mathfrak{n}}+1) \mathcal{D}[\hat a]\hat\rho+\kappa\bar{\mathfrak{n}} \mathcal{D}[\hat a^\dagger]\hat\rho,
\end{equation}
where $\hat a$ represents the bosonic mode annihilation operator, and the Lindblad superoperator is given by
\begin{equation}
    \mathcal{D}[\hat O]\hat\rho=2\hat O\hat\rho\hat O^\dagger-\hat\rho\hat O^\dagger\hat O-\hat O^\dagger\hat O\hat\rho .
\end{equation}
For the Hamiltonian, we assume a generalized form of the Dicke model,
\begin{equation}
\begin{split}
	\hat H&=\omega\hat a^\dagger \hat a+\omega_0\hat S_z+\frac{\lambda_-}{\sqrt{2S}}\left(\hat a\hat S_++ \hat a^\dagger\hat S_-\right)\\
	&+\frac{\lambda_+}{\sqrt{2S}}\left(\hat a\hat S_-+ \hat a^\dagger\hat S_+\right) .
\end{split}
\end{equation}
This is expressed in terms of the collective spin operators which are sums of the spin operators of the individual spinors $\hat S^{(n)}_{\{x,y,z,\pm\}}$
\begin{equation}
    \hat S_{\{x,y,z,\pm\}}=\sum_n^N \hat S^{(n)}_{\{x,y,z,\pm\}}.
\end{equation}

In the dispersive limit ($|\omega|\gg\omega_0,\lambda_\pm$, albeit more generally valid for $\sqrt{\omega^2+\kappa^2}\gg\omega_0,\lambda_\pm$) we can adiabatically eliminate the bosonic mode to obtain 
a master equation for the reduced density operator of the collective spin in the form (see Appendix A) and \cite{PhysRevLett.119.213601}
\begin{equation}
\label{reducedmasterequation}
\begin{split}
    \dot{\hat\rho}&=-i\left[\hat H,\hat\rho\right]+\frac{\kappa(\bar{\mathfrak{n}}+1)}{2S(\omega^2+\kappa^2)}\mathcal{D}[\lambda_-\hat S_-+\lambda_+\hat S_+]\hat\rho\\
    &+\frac{\kappa\bar{\mathfrak{n}}}{2S(\omega^2+\kappa^2)}\mathcal{D}[\lambda_-\hat S_++\lambda_+\hat S_-]\hat\rho ,
\end{split}
\end{equation}
with the Hamiltonian
\begin{equation}
\begin{split}
\label{Hamiltonianadbelim}
\hat H&=\left[\omega_0-\frac{\omega(2\bar{\mathfrak{n}}+1)(\lambda^2_--\lambda^2_+)}{2S(\omega^2+\kappa^2)}\right]\hat S_z\\
&-\frac{\omega}{2S(\omega^2+\kappa^2)}\left[(\lambda_-+\lambda_+)^2\hat S_x^2+(\lambda_--\lambda_+)^2\hat S_y^2\right].
\end{split}
\end{equation}

Now, for the purposes of this paper, the original Dicke model system is itself assumed to have been ``engineered'' and that the parameters $\omega$, $\omega_0$, and $\lambda_\pm$ are effective parameters that are typically comparable in (energy) scale, but can be tuned as desired. So, in particular, we may choose 
$\lambda_+=\lambda_-=\lambda$ and $\omega_0=0$ in Eqs.~(\ref{reducedmasterequation}-\ref{Hamiltonianadbelim}) to give the one-axis twisting Hamiltonian \cite{PhysRevA.47.5138},
\begin{equation}
\label{hamiltonian}
	\hat H=-\Lambda\hat S_x^2,
\end{equation}
with the master equation
\begin{equation}\label{ME}
    \dot{\hat\rho}=-i\left[\hat H,\hat\rho\right]+\Gamma\mathcal{D}[\hat S_x]\hat\rho,
\end{equation}
where
\begin{equation}
    \Lambda=\frac{2\lambda^2\omega}{S(\omega^2+\kappa^2)}
\end{equation}
and
\begin{equation}
    \Gamma=\frac{\kappa(2\bar{\mathfrak{n}}+1)}{\omega}\Lambda.
\end{equation}
For the remainder of this paper $\omega$ and $\Lambda$ are assumed, without loss of generality, to be positive.

\subsection{Time evolution of the wave function}
\label{basischange}
The one-axis twisting Hamiltonian (\ref{hamiltonian}) is known to generate spin cat states, that is, coherent superpositions of the angular momentum eigenstates $|S,S\rangle_z$ and $|S,-S\rangle_z$. 
For finite $\Gamma$, we can consider a quantum trajectory treatment of (\ref{ME}) and generalize this Hamiltonian evolution to that of the non-Hermitian effective Hamiltonian \cite{carmichael1993open,Molmer:93}
\begin{equation}
\label{nonhermH}
	\mathcal{\hat H}=-(\Lambda+i\Gamma)\hat S_x^2.
\end{equation}
In the two systems that we will be looking at, it turns out that either $\Gamma\approx0$ (trapped ions) or that, for finite $\Gamma$ (cavity QED), the quantum jumps associated with the quantum trajectory picture correspond to photon emissions from a cavity, which could be monitored so as to facilitate a conditional evolution. 
We note also the possibility $\Lambda=0$ and $\Gamma >0$. In this case, the Hamiltonian can lead to novel entangled-state cycles of collective spin states or to probabilistic preparation of the state $|S,0\rangle_x$ \cite{PhysRevA.77.033810}.

Since the Hamiltonian (\ref{nonhermH}) depends only on $\hat S_x$, it will be convenient at times to change from the basis of $\hat S_z$ eigenstates to the basis spanned by those of $\hat S_x$ through
\begin{equation}
	|S,m'\rangle_z=\sum_{m=-S}^{S}D_{m,m'}^S\left(-\frac{\pi}{2}\right)|S,m\rangle_x ,
\end{equation}
where the subscript on the ket indicates in which basis the collective spin state is, and $D_{m,m'}^S$ are the elements of the Wigner $D$ matrix \cite{sakurai2017modern},
\begin{equation}
\begin{split}
D^S_{m,m'}(\beta)&=\sum_{k=0}^{S+m}(-1)^{k-m'+m}\\
&\times\frac{\sqrt{(S+m')!(S-m')!(S+m)!(S-m)!}}{k!(S+m'-k)!(S-m-k)!(k+m-m')!}\\
&\times\left(\cos\frac{\beta}{2}\right)^{2S-2k+m'-m}\left(\sin\frac{\beta}{2}\right)^{2k-m'+m} ,
\end{split}
\end{equation}
where the sum runs over all non-negative factorial arguments. 
From now on we can simplify the notation, as we always consider the case $\beta=-\frac{\pi}{2}$ and all sums over $m$ are from $-S$ to $S$. In particular (from the representation as a spin coherent state \cite{PhysRevA.6.2211}) we have
\begin{equation}
    D^S_{m,S}(-\frac{\pi}{2})=(-1)^{S-m}2^{-S}\sqrt{\binom{2S}{S+m}}.
\end{equation}
We always start with an initial Dicke state with maximal orientation in $z$-direction,
\begin{equation}
|\psi(0)\rangle=|S,S\rangle_z.
\end{equation}
During the evolution with (\ref{nonhermH}) the norm of the wave function decays due to the non-Hermiticity; the renormalized wave function at a time $t$ is given by
\begin{equation}
\begin{split}
\label{wavefunction}
	|\psi(t)\rangle&=\frac{e^{-i\mathcal{\hat H}t}|\psi(0)\rangle}{||e^{-i\mathcal{\hat H}t}|\psi(0)\rangle||}\\
	&=\frac{\sum_m D^S_{m,S}e^{i\Lambda m^2 t}e^{-\Gamma m^2t}|S,m\rangle_x}{\sqrt{\sum_m e^{-2\Gamma m^2t}(D^S_{m,S})^2}}.
\end{split}
\end{equation}

\subsection{Fidelity and Quantum Fisher Information}
Throughout this paper we will use the fidelity, $F$, and Quantum Fisher Information (QFI) \cite{PhysRevLett.72.3439} (with respect to the generator $\hat S_z$), $\mathcal{F}[\hat S_z]$, as figures of merit. The former is a metric for the closeness of the actual quantum state, as described by the density operator $\hat\rho$, to the target state $|\Psi\rangle$, and the latter for its quantum metrological usefulness. They are computed in their most general form through
\begin{equation}
    F=\left(\text{Tr}\sqrt{\sqrt{|\Psi\rangle\langle\Psi|}\hat\rho\sqrt{|\Psi\rangle\langle\Psi|}}\right)^2
\end{equation}
and
\begin{equation}
    \mathcal{F}[\hat S_z]=2\sum_{k,k'}\frac{(\lambda_k-\lambda_{k'})^2}{\lambda_k+\lambda_{k'}}\left|\langle e_k|\hat S_z|e_{k'}\rangle\right|^2,
\end{equation}
where $\lambda_k$ and $|e_k\rangle$ are the eigenvalues and eigenvectors of $\hat\rho$, respectively. In the case of a pure state $\hat\rho =|\psi\rangle\langle\psi |$ these simplify to
\begin{equation}
    F=|\langle\Psi|\psi\rangle|^2
\end{equation}
and
\begin{equation}
    \mathcal{F}[\hat S_z]=4\left(\langle\hat S_z^2\rangle-|\langle\hat S_z\rangle|^2\right).
\end{equation}


\subsection{No-jump trajectory: Cat State}
\label{nojumps}
Free evolution of the initial state subject to the effective Hamiltonian (\ref{nonhermH}) up to a time $ t=\pi /(2\Lambda )$ gives the state (see Appendix B)
\begin{equation}
\begin{split}
	|\psi(t=\frac{\pi}{2\Lambda})\rangle
		&=\frac{\frac{1+i}{2}\sum_m D^S_{m,S}e^{-\frac{\pi\Gamma}{2\Lambda} m^2}|S,m\rangle_x}{\sqrt{\sum_m e^{-\frac{\pi\Gamma}{\Lambda} m^2}(D^S_{m,S})^2}}\\
	&+\frac{(-1)^S\frac{1-i}{2}\sum_m D^S_{m,-S}e^{-\frac{\pi\Gamma}{2\Lambda} m^2}|S,m\rangle_x}{\sqrt{\sum_m e^{-\frac{\pi\Gamma}{\Lambda} m^2}(D^S_{m,S})^2}}.
\end{split}
\end{equation}
In the case of vanishing decay ($\Gamma=0$) this reduces to
\begin{equation}
	\begin{split}
	|\Psi\rangle&=
	\frac{1+i}{2}\sum_m D^S_{m,S}|S,m\rangle_x\\
	&+(-1)^S\frac{1-i}{2}\sum_m D^S_{m,-S}|S,m\rangle_x\\
	&=\frac{1+i}{2}|S,S\rangle_z+(-1)^S\frac{1-i}{2}|S,-S\rangle_z ,
	\end{split}
\end{equation}
i.e., a spin cat state. More generally, the cat state occurs at times $\Lambda t=\frac{\pi}{2} \mod 2\pi$, while at times $\Lambda t=\frac{3\pi}{2} \mod 2\pi$ one obtains another spin cat state with a different relative phase,
\begin{equation}
    |\Psi'\rangle=\frac{1+i}{2}|S,S\rangle_z+(-1)^{S+1}\frac{1-i}{2}|S,-S\rangle_z.
\end{equation}
 Note that if the initial state is instead $|S,-S\rangle_z$, then one obtains $|\Psi'\rangle$ at $\Lambda t=\frac{\pi}{2} \mod 2\pi$.

From the overlap between $|\psi(t)\rangle$ and the target state $|\Psi\rangle$ at time $t=\pi /(2\Lambda )$,
\begin{equation}
	\begin{split}
	\langle\Psi|\psi( t=\frac{\pi}{2\Lambda})\rangle
	&=\frac{\sum_m e^{-\frac{\pi\Gamma}{2\Lambda} m^2}(D^S_{m,S})^2}{\sqrt{\sum_m e^{-\frac{\pi\Gamma}{\Lambda} m^2}(D^S_{m,S})^2}},
	\end{split}
\end{equation}
we obtain the fidelity as
\begin{equation}
F=\frac{\left[\sum_m e^{-\frac{\pi\Gamma}{2\Lambda} m^2}(D^S_{m,S})^2\right]^2}{\sum_m e^{-\frac{\pi\Gamma}{\Lambda} m^2}(D^S_{m,S})^2},
\end{equation}
while the QFI is computed to be
\begin{equation}
\begin{split}
    &\mathcal{F}[\hat S_z]=4\frac{\sum_{m,k,l} D^S_{m,S}D^S_{k,S}D^S_{m,l}D^S_{k,l}e^{-\frac{\pi\Gamma}{\Lambda} (m^2+k^2)}i^{k^2-m^2} l^2}{\sum_m e^{-\frac{\pi\Gamma}{\Lambda} m^2}(D^S_{m,S})^2}\\
    &-4\left|\frac{\sum_{m,k,l} D^S_{m,S}D^S_{k,S}D^S_{m,l}D^S_{k,l}e^{-\frac{\pi\Gamma}{2\Lambda} (m^2+k^2)}i^{k^2-m^2} l}{\sum_m e^{-\frac{\pi\Gamma}{\Lambda} m^2}(D^S_{m,S})^2}\right|^2.
\end{split}
\end{equation}



\begin{figure}[t]
\centering
	\includegraphics[width=0.49\linewidth]{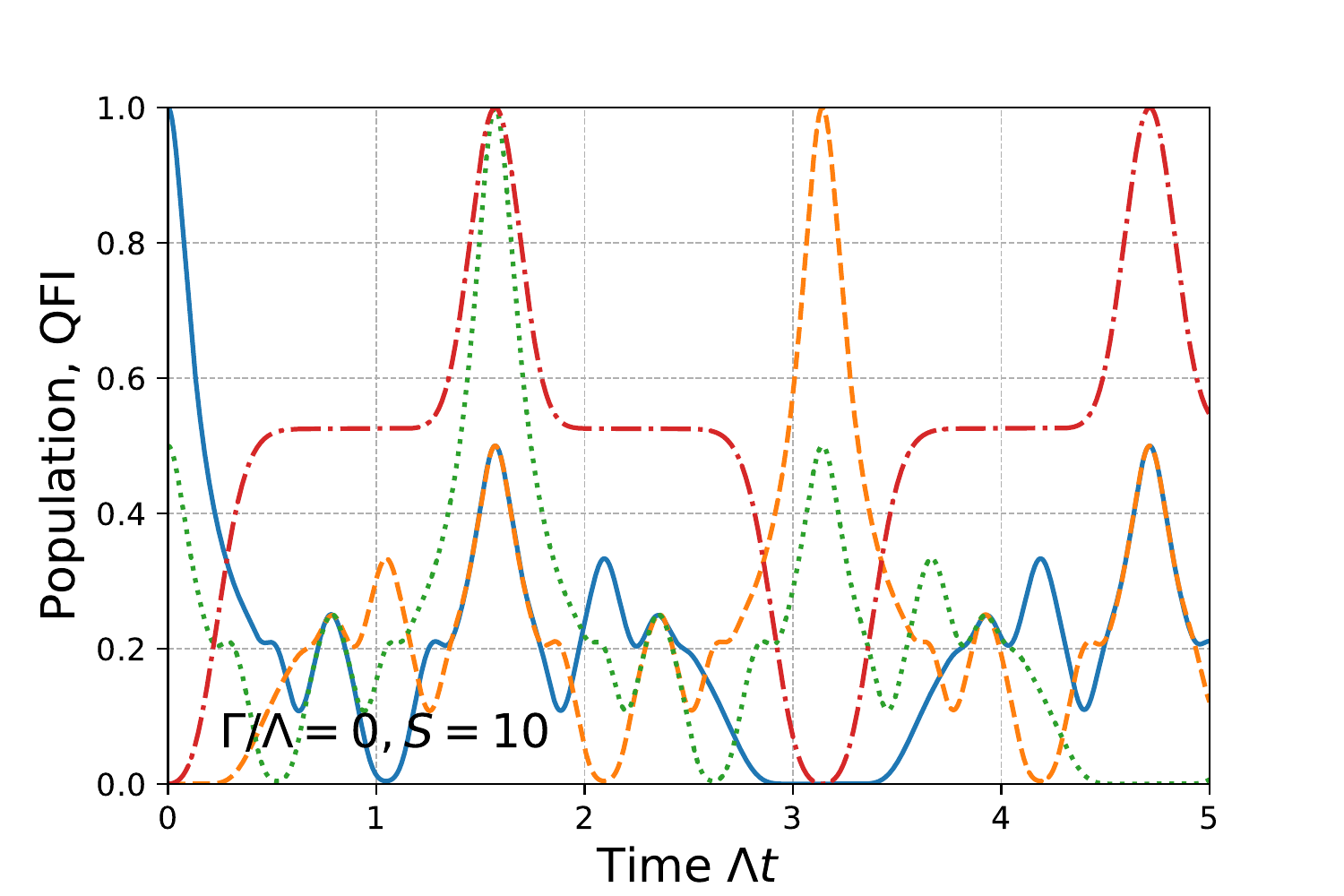}
	\includegraphics[width=0.49\linewidth]{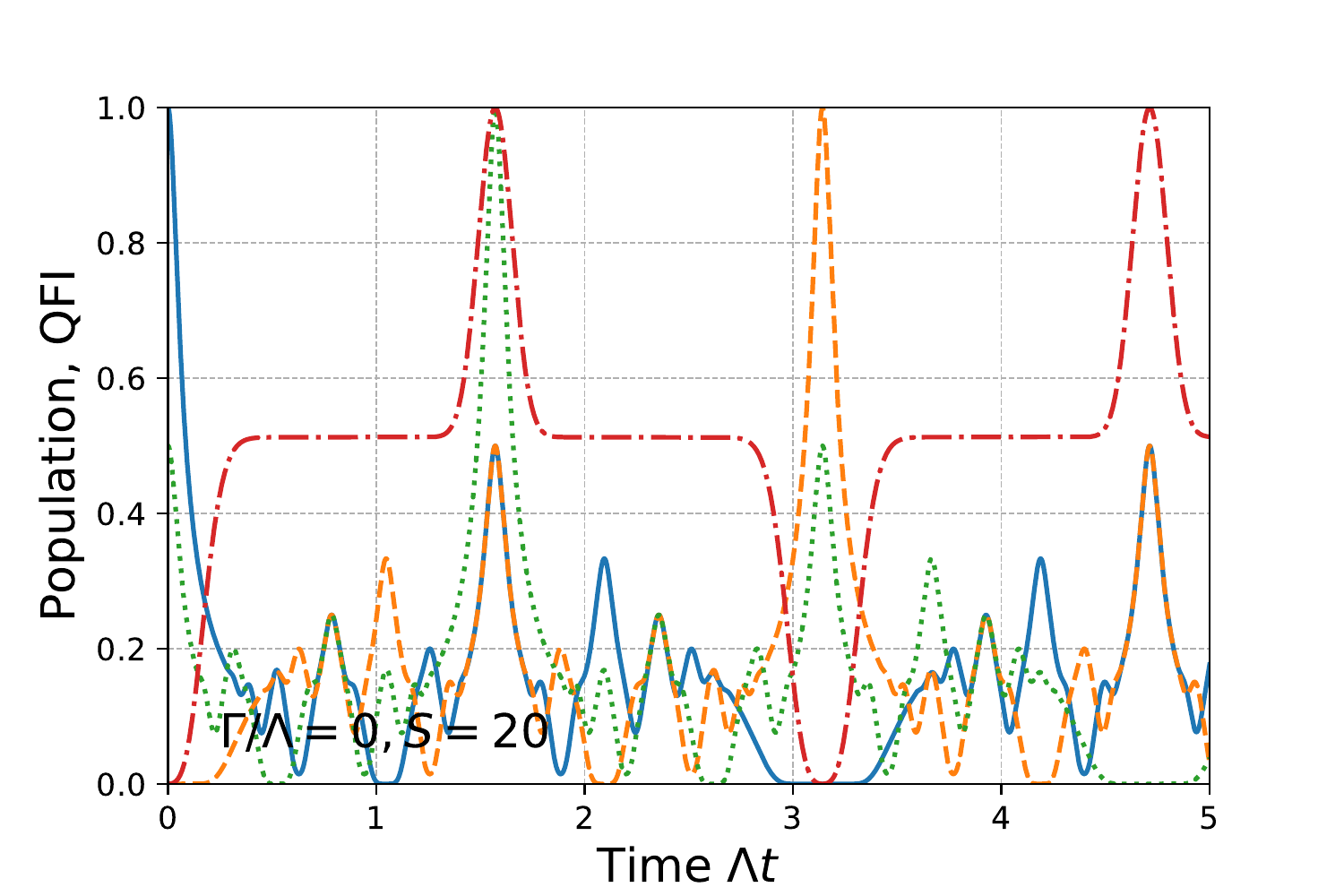}\\
	\includegraphics[width=0.49\linewidth]{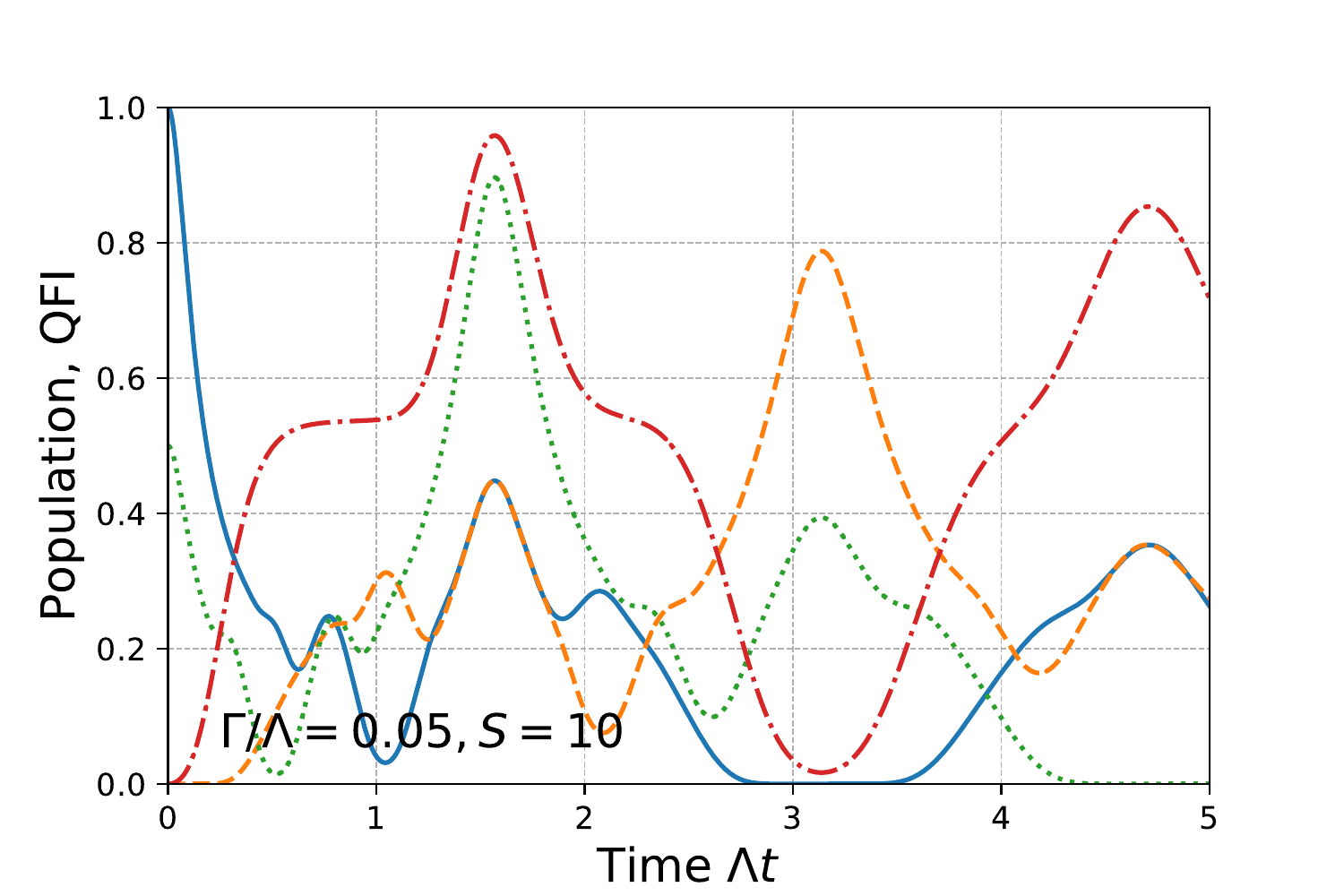}
	\includegraphics[width=0.49\linewidth]{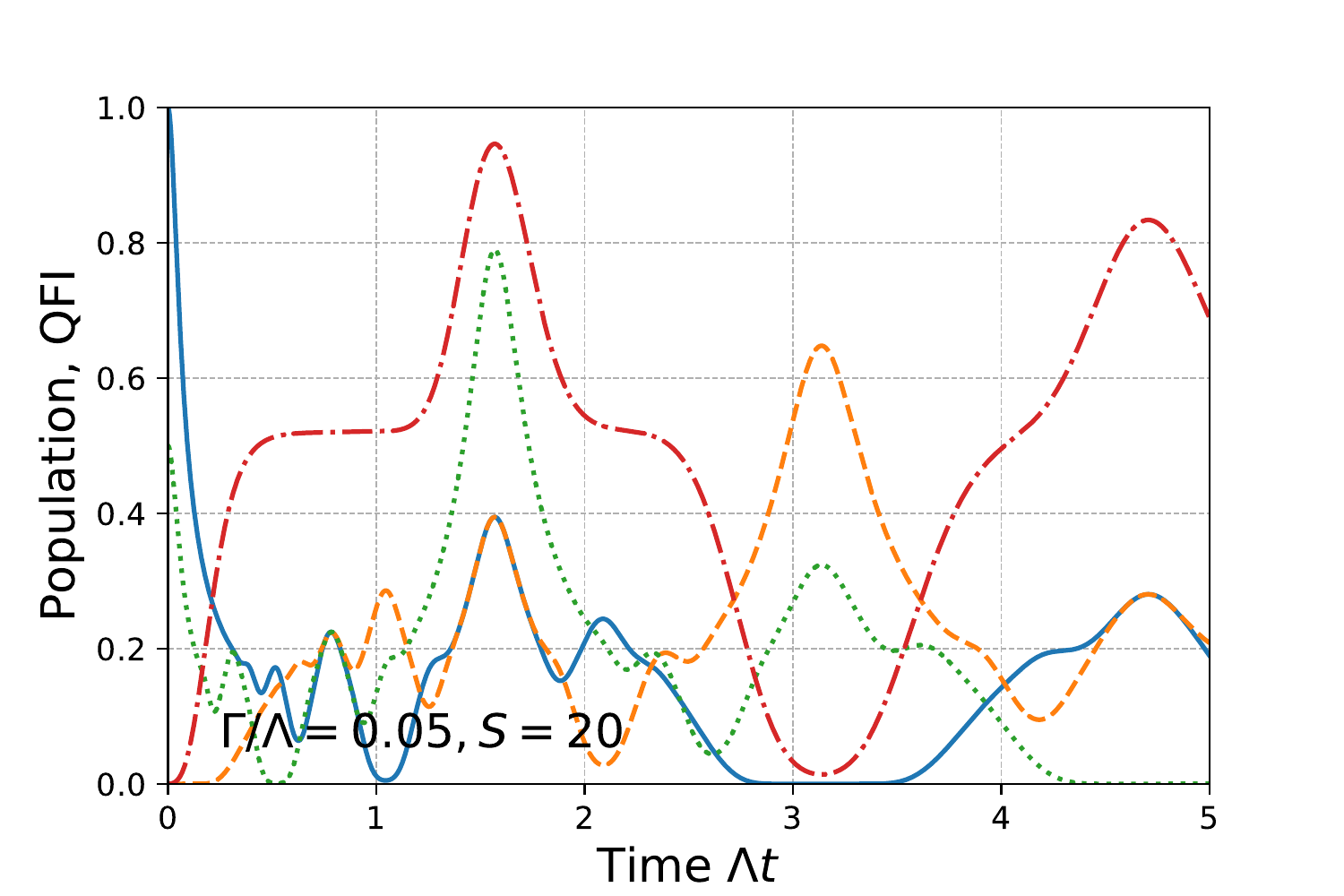}
	\caption{Master equation simulations of the QFI (red dash-dotted) and the populations of $|S,S\rangle_z$ (blue solid), $|S,-S\rangle_z$ (orange dashed) and $|\Psi\rangle $ (green dotted) over time for spin $S=10$ (left column) and $S=20$ (right column), with $\frac{\Gamma}{\Lambda}=0$ (top row) and $\frac{\Gamma}{\Lambda}=0.05$ (bottom row). The QFI is plotted relative to the theoretical maximum, $4S^2$.}
	\label{poptemp}
\end{figure}

In Fig. \ref{poptemp} we show the temporal evolution of the populations of the various states of interest, i.e., the states $|S,S\rangle_z$ and $|S,-S\rangle_z$, and the target state $|\Psi\rangle$, for two values of the total spin $S$ and for $\Gamma=0$ and $\Gamma>0$. It is clear that a higher spin makes the system more fragile to finite dissipation, although more so with regards to the fidelity than to the QFI (we will look at this in more detail in Section IV). Note that the revival of the QFI at $\Lambda t=\frac{3\pi}{2}$ corresponds to generation of the state $|\Psi'\rangle$.

\section{Trapped-Ion Framework}
\label{ions}

\subsection{Implementation}
For this framework we follow the proposal of \cite{PhysRevA.97.042317} in which the spinors are two-level ions, with transition frequency $\omega_\text{TS}$, confined in a trap of frequency $\nu$, and the harmonic oscillator mode corresponds to the center-of-mass vibrational mode of the ions. The ions are driven by two lasers, one red-detuned and one blue-detuned from the transition frequency, with bare frequencies $\omega_\pm$ and Rabi frequencies $\Omega_\pm$. In the Lamb-Dicke regime, i.e., $\eta\sqrt{2\bar{\mathfrak{n}}+1}\ll1$, where $\eta$ is the Lamb-Dicke parameter and $\bar{\mathfrak{n}}$ is the mean phonon number in the center-of-mass mode, it can be shown that the system is described by an effective Dicke model with parameters given by \cite{PhysRevA.97.042317}
\begin{equation}
    \begin{split}
    \omega&=\nu+\frac{\omega_--\omega_+}{2},\\
    \omega_0&=\omega_\text{TS}-\frac{\omega_-+\omega_+}{2},\\
    \lambda_\pm&=\frac{\sqrt{2S}\,\Omega_\pm\eta}{2}.
    \end{split}
\end{equation}
For our purposes, we require that $\omega_\pm=\omega_\text{TS}\pm\nu+\delta_\pm$ with the offsets $\delta_\pm$ chosen to satisfy $\delta_+=-\delta_-=-\delta$, so that $\omega=\delta$ and $\omega_0=0$. We also assume that $\Omega_+=\Omega_-=\Omega$.

\subsection{Damping and Dephasing}
Heating rates in ion traps are typically very small, while advanced cooling schemes (for example, using electromagnetically induced transparency \cite{PhysRevA.93.053401}) are expected to enable cooling close to the ground state, corresponding to $\bar{\mathfrak{n}}\approx 0$. Hence, within the trapped-ion framework it is reasonable to assume that $\Gamma\simeq 0$ and to neglect this source of damping.

The dominating decoherence process is usually dephasing originating from voltage fluctuations that propagate to the trap frequency, magnetic field and laser frequency \cite{haffner2008quantum}. We use the following Master equation
\begin{equation}
\label{dephasingME}
    \dot{\hat\rho}=-i\left[\hat H,\hat\rho\right]+\frac{\epsilon}{2}\sum_n^N\mathcal{D}[\hat S^{(n)}_z]\hat\rho,
\end{equation}
where $\epsilon$ is the dephasing rate.

\begin{figure}[t]
\centering
	\includegraphics[width=\linewidth]{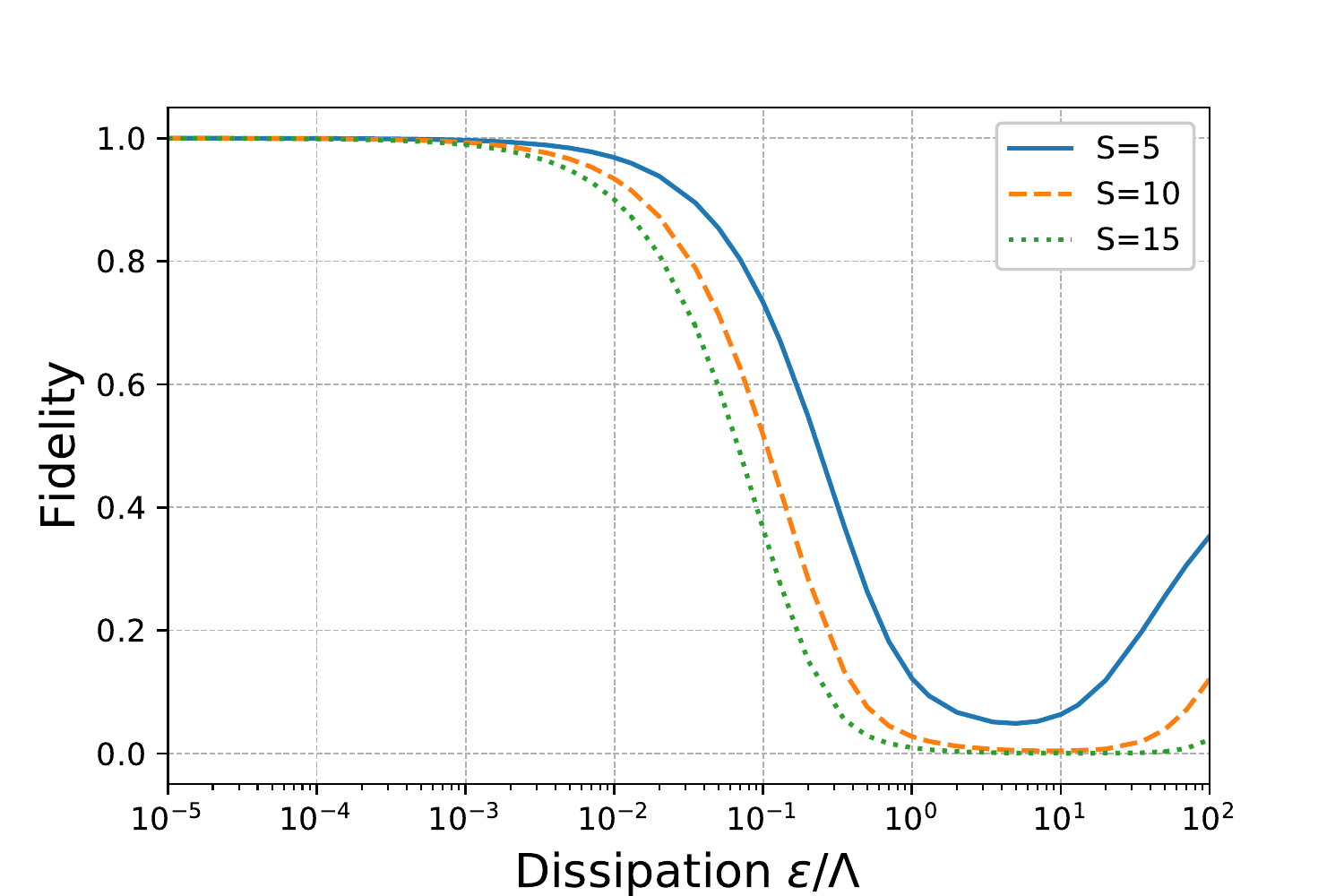}\\
	\includegraphics[width=\linewidth]{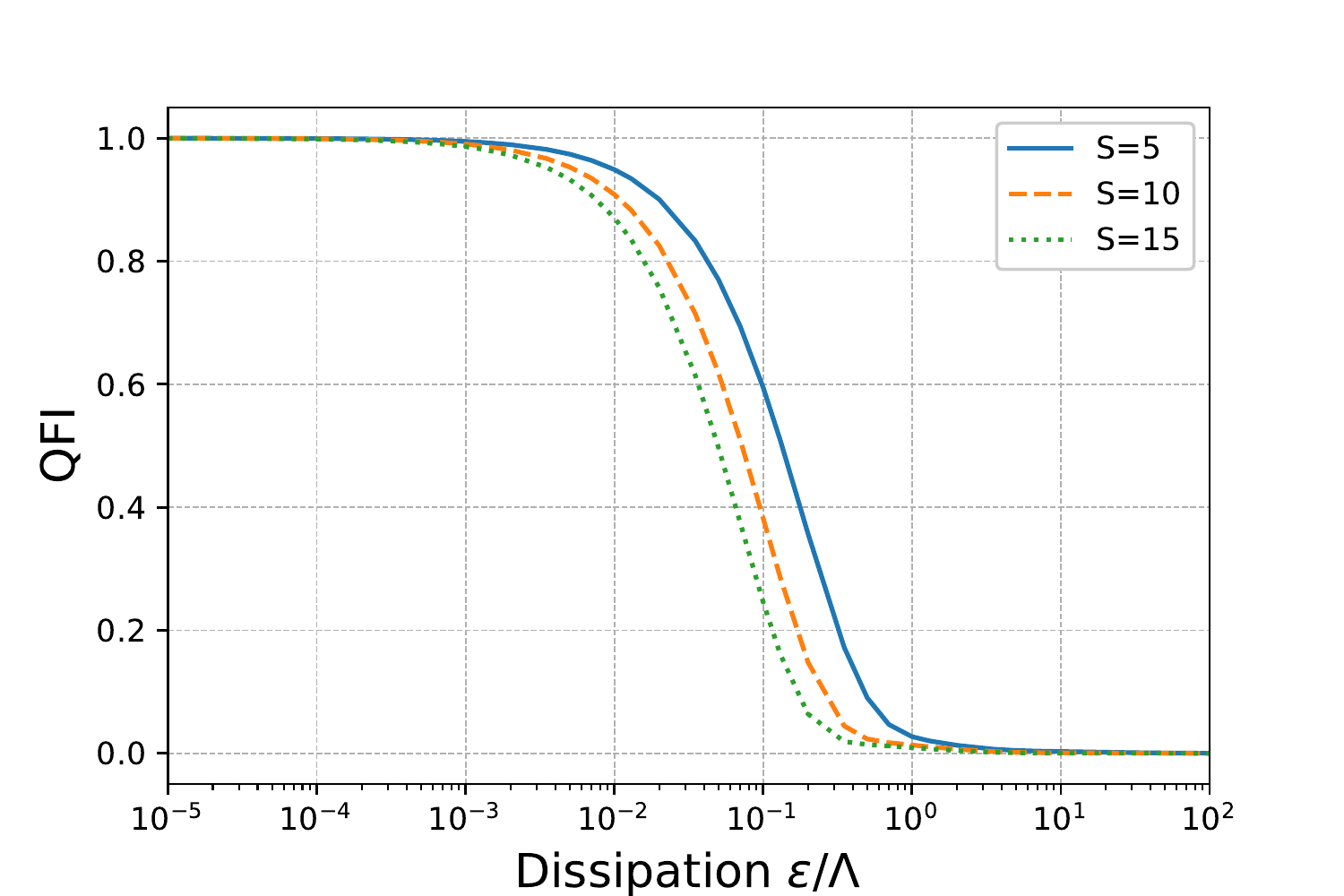}
	\caption{Fidelity (top) and QFI (bottom) as a function of the strength of dissipation $\frac{\epsilon}{\Lambda}$ at the time $\Lambda t=\frac{\pi}{2}$. The QFI is plotted relative to the theoretical maximum, $4S^2$.}
	\label{dephasing}
\end{figure}

The dynamics in (\ref{dephasingME}) are invariant under permutation of the identical ions and can therefore be modeled in a Dicke state basis \cite{zhang2018monte,PIQS}. Hence, we use the Permutational Invariant Quantum Solver (PIQS) \cite{PIQS} to solve the master equation.
Fig.~\ref{dephasing} plots the fidelity and the QFI as a function of the dephasing rate $\epsilon$. Again the overall trend is a decrease in these two quantities, however, in the limit of very strong dephasing the system essentially gets trapped in the state $|S,S\rangle_z$ and the larger the dephasing the slower the leakage out of that state. This explains why we see the fidelity go up again for increasing dephasing rate $\epsilon$, as the overlap between the catstate and $|S,S\rangle_z$ is large (the fidelity goes to $|\langle\Psi|\psi\rangle|^2=\frac{1}{2}$ in the limit of infinite $\epsilon$), even though the QFI goes to zero.

\subsection{Possible Experimental Parameters}

For a specific experimental configuration, we can consider trapped $^{40}$Ca$^+$ ions that are driven on the $S_{1/2}\xleftrightarrow{} D_{5/2}$ quadrupole transition \cite{gerritsma2010quantum}. This transition has a very small natural linewidth, which allows us to ignore spontaneous emission in what follows, especially since the lasers are also detuned from the transition by roughly the trap frequency ($\Delta=\omega_\pm-\omega_{TS}=\pm\nu\mp\delta$). 

Let us consider an ensemble of $N=20$ ions ($S=10$) in a trap of frequency $\nu/2\pi=3$~MHz and Lamb-Dicke parameter $\eta=0.05$. With $\Omega/2\pi=300$~kHz, we have $\lambda/2\pi=34$~kHz. The effective coupling strength in the one-axis twisting Hamiltonian is given, for the proposed ion system, by
\begin{equation}
    \Lambda=\frac{\Omega^2\eta^2}{\delta},
\end{equation}
so selecting $\delta/2\pi=300$~kHz ($\gg\lambda$) yields $\Lambda/2\pi=750$~Hz. Looking at Fig.~\ref{dephasing}, we see that to achieve a fidelity in excess of 0.9 requires a dephasing rate $\epsilon/\Lambda\lesssim 0.02$, or $1/\epsilon\gtrsim 11$~ms, which appears within reach of recent experiments with strings of $^{40}$Ca$^+$ ions \cite{PhysRevLett.124.240505}.



\section{Cavity QED Framework}
\label{cavityqed}
\subsection{Implementation}
\begin{figure}[t]
	\includegraphics[width=0.49\linewidth]{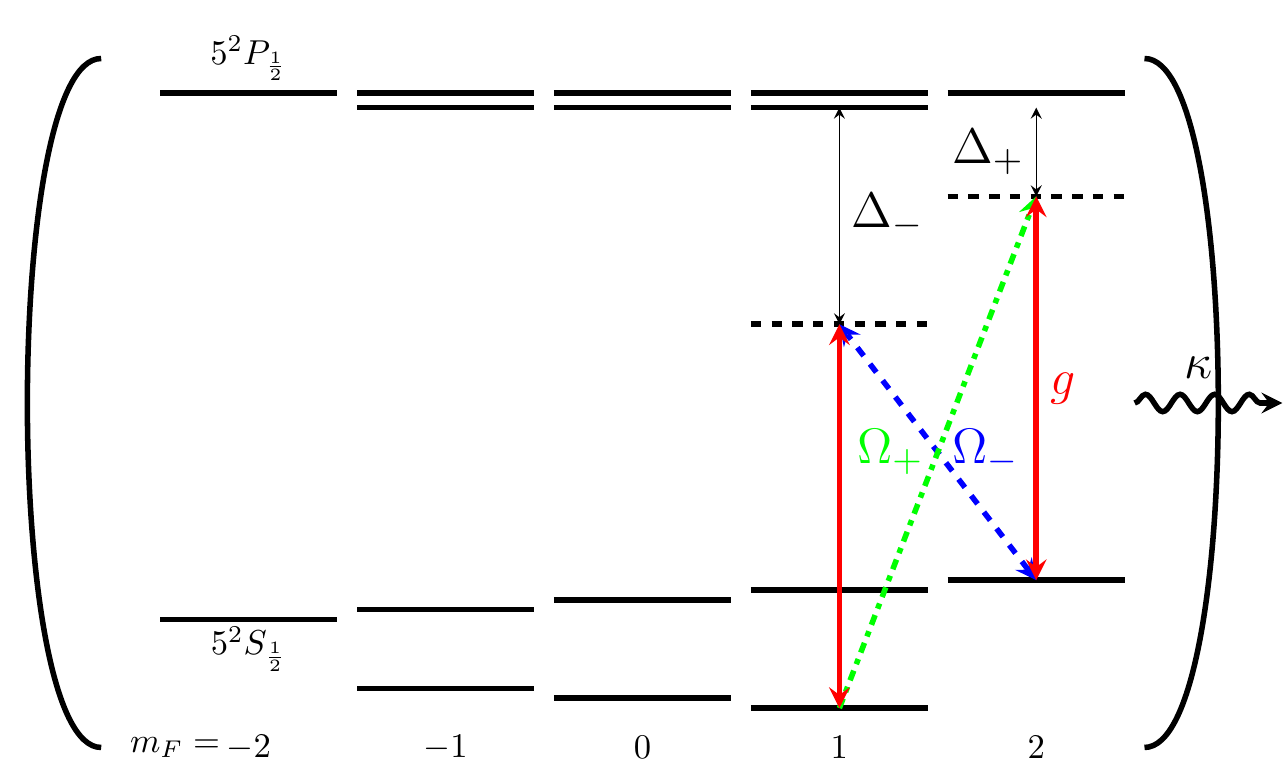}
	\includegraphics[width=0.49\linewidth]{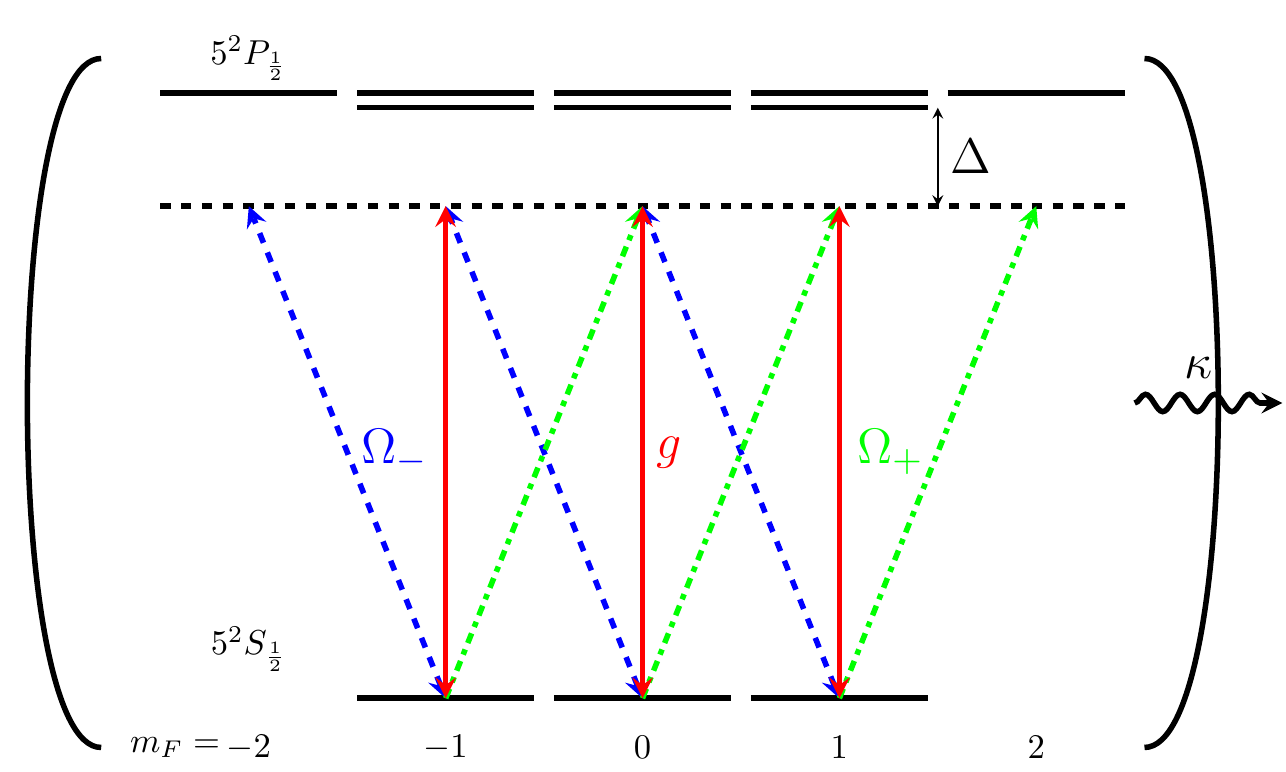}
	\caption{Atomic level configurations and excitation schemes for spin-$\frac{1}{2}$ (left) and spin-$1$ (right) realizations of the Dicke model in optical cavity QED with $^{87}$Rb atoms.}
	\label{scheme}
\end{figure}

For this framework we consider an ensemble of atoms trapped inside an optical cavity and undergoing Raman transitions between selected hyperfine-ground-state levels. The Raman transitions are driven by the cavity field and auxiliary laser fields, as depicted in Fig.~\ref{scheme}. Such schemes have been used in experimental realizations of the Dicke model and implement either effective spin-1/2 \cite{PhysRevA.97.043858} or spin-1 \cite{Zhiqiang:17} atoms. Note that using the scheme of \cite{Zhiqiang:17} one could also implement larger effective spins per atom by using other hyperfine levels (e.g., spin-2 in $^{87}$Rb) or atomic species (e.g., spin-3 or spin-4 for $^{133}$Cs).
Note also that the spin-1/2 implementation is also feasible using the clock states (rather than the stretched states, as depicted in Fig.~\ref{scheme}) of $^{87}$Rb or $^{133}$Cs \cite{PhysRevLett.110.120402}. However, for the spin-1/2 realization an additional term, proportional to $\hat S_z\hat a^\dagger\hat a$, appears in the effective Hamiltonian (the effect of which, however, becomes small for large enough detunings of the fields), and the tuning of laser frequencies and driving strengths requires somewhat more care.

For the spin-1 realization in $^{87}$Rb the parameters of the effective Dicke model take the specific forms \cite{PhysRevLett.119.213601}
\begin{equation}
\begin{split}
	\omega&=\omega_c-\frac{\omega_-+\omega_+}{2} +\frac{Sg^2}{3\Delta},\\
	\omega_0&=\omega_z-\frac{\omega_--\omega_+}{2} +\frac{\Omega_-^2-\Omega_+^2}{24\Delta},\\
	\lambda_\pm&=\frac{\sqrt{S}g\Omega_\pm}{12\Delta},\\
\end{split}
\end{equation}
where $\omega_{\rm c}$ and $\omega_\pm$ are the frequencies of the cavity mode and $\sigma_\pm$-polarized laser field, respectively, $\Omega_\pm$ the Rabi frequencies of the laser fields, $g$ the single-atom-cavity coupling strength, $\omega_{\rm z}$ the Zeeman splitting of the $F=1$ atomic levels, and $\Delta$ the detuning of the lasers and cavity mode from the excited state manifold.

Such a spin-1 implementation has been realized in the dispersive regime in \cite{PhysRevLett.122.010405}.

\subsection{Cavity Decay: No-Jump Evolution}
\begin{figure}[b]
\centering
	\includegraphics[width=\linewidth]{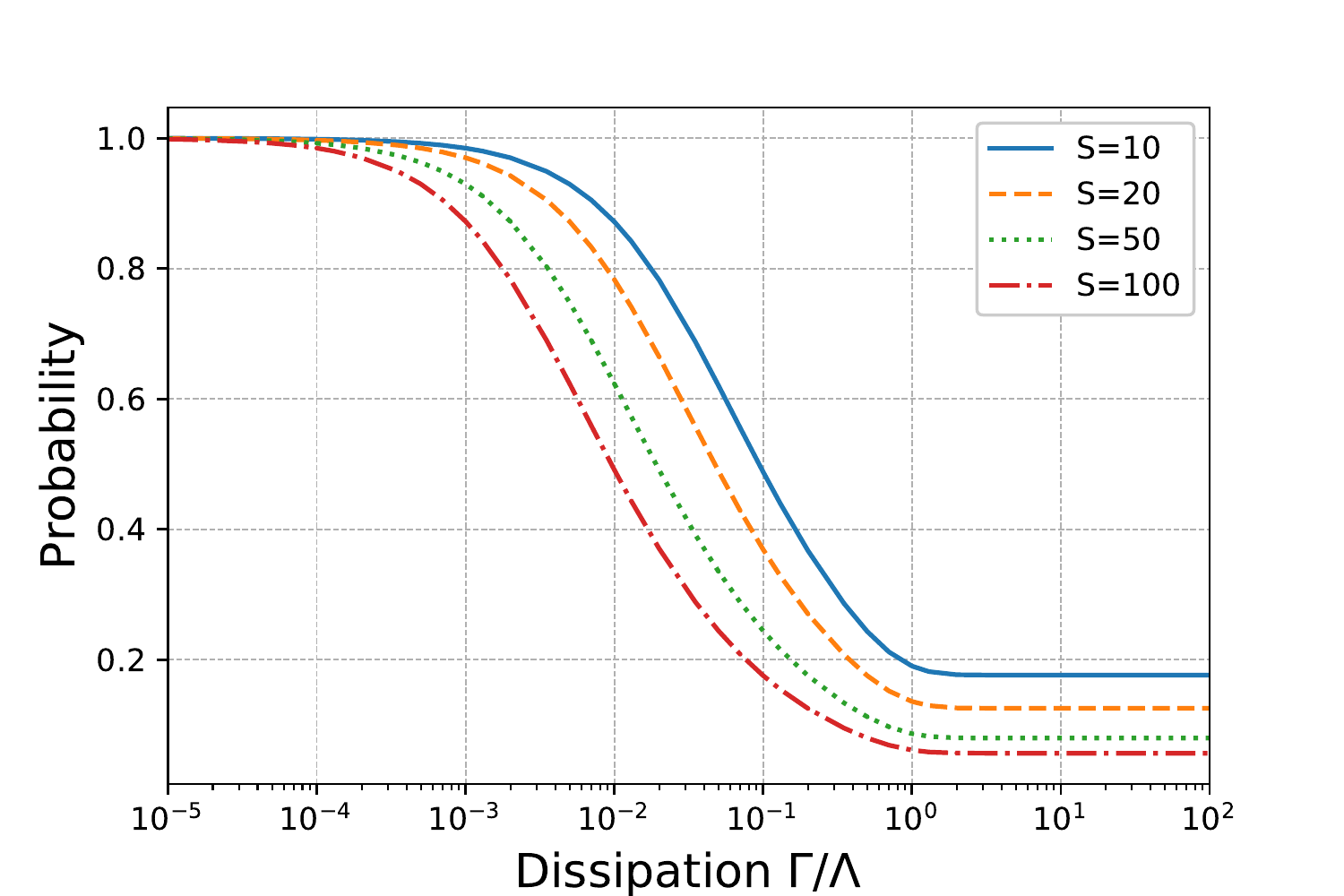}
	\caption{Probability of there being no jump until a time $t=\pi /(2\Lambda )$ as a function of the dissipation $\Gamma /\Lambda$ for a selection of different spin lengths.}
	\label{prob}
\end{figure}

The normalization factor in Eq.~(\ref{wavefunction}),
\begin{equation}
||e^{-i\mathcal{\hat H}t}|\psi(0)\rangle||=\sqrt{\sum_m e^{-2\Gamma m^2t}(D^S_{m,S})^2} ,
\end{equation}
gives us the probability $P(t)$ that no jump (no photon emission) has occurred up to a time $t$ through
\begin{equation}
P(t)=||e^{-i\mathcal{\hat H}t}|\psi(0)\rangle||^2=\sum_m e^{-2\Gamma m^2t}(D^S_{m,S})^2.
\end{equation}
Since the occurrence of a jump or not leads to two very distinct behaviours, it makes sense to take a closer look at the probability of there being no jump up until a time $t=\pi/(2\Lambda )$, 
\begin{equation}
P(t=\frac{\pi}{2\Lambda})=\sum_m e^{-\frac{\pi\Gamma}{\Lambda} m^2}(D^S_{m,S})^2 ,
\end{equation}
for reasons that will become evident in the next section. 
A certain level of dissipation can be tolerated without considerably increasing the possibility of a jump, but that level clearly decreases with increasing total spin, as shown in Fig.~\ref{prob}, where $P(t=\pi /(2\Lambda ))$ is plotted as a function of $\Gamma /\Lambda$ for varying values of $S$.

Note that if no jump has occurred after a sufficiently long time (i.e., $2\Gamma t\gg 1$), then the system will be projected into the state $|S,0\rangle_x$, as this is the only component for which the probability amplitude does not decay. This explains why, in Fig.~\ref{prob}, the probability curves level out at constant, finite values for $\Gamma /\Lambda\gtrsim 1$. However, the likelihood of there being no jump decreases gradually with increasing spin length \cite{PhysRevA.77.033810}.

\begin{figure}[t]
\centering
	\includegraphics[width=\linewidth]{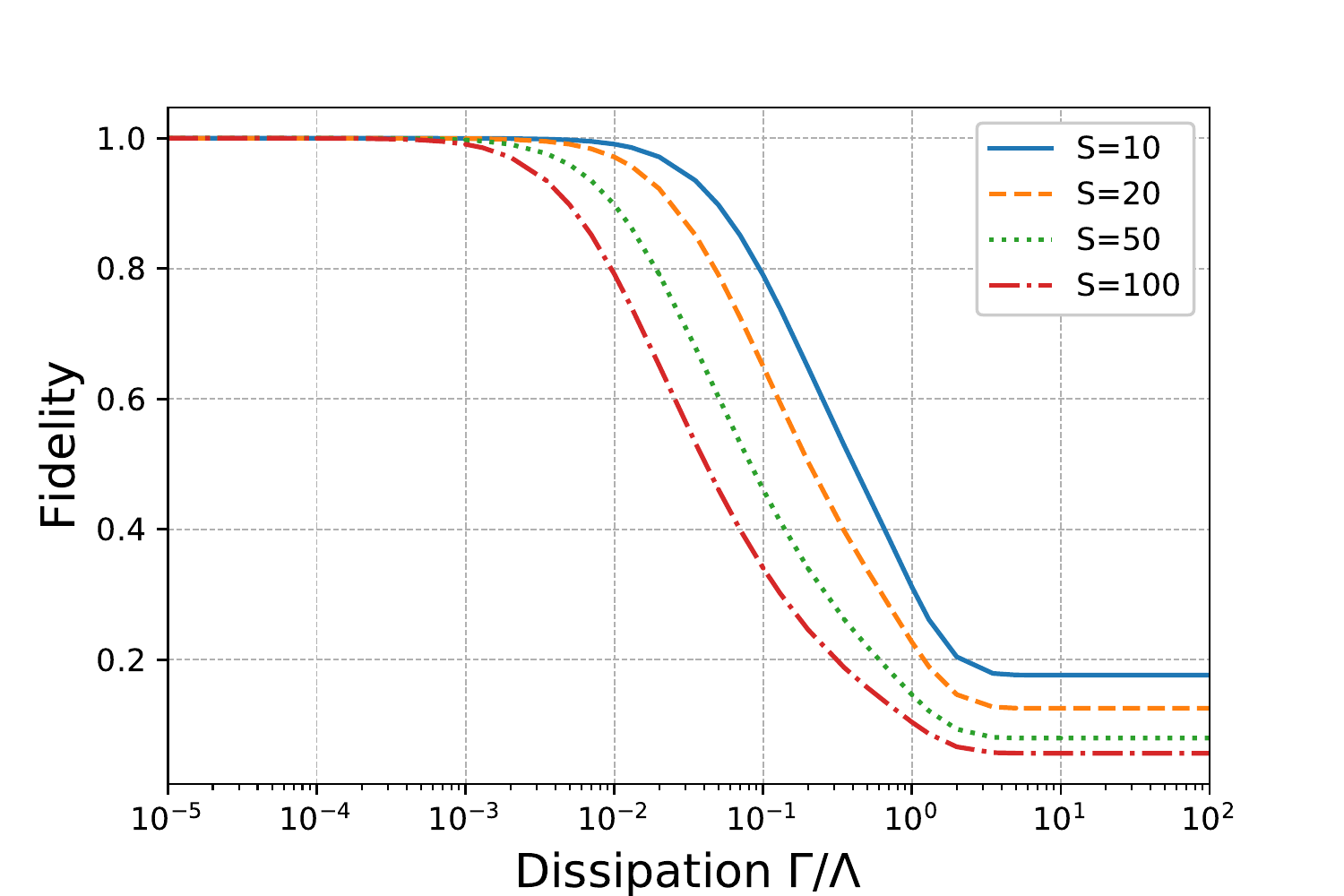}\\
	\includegraphics[width=\linewidth]{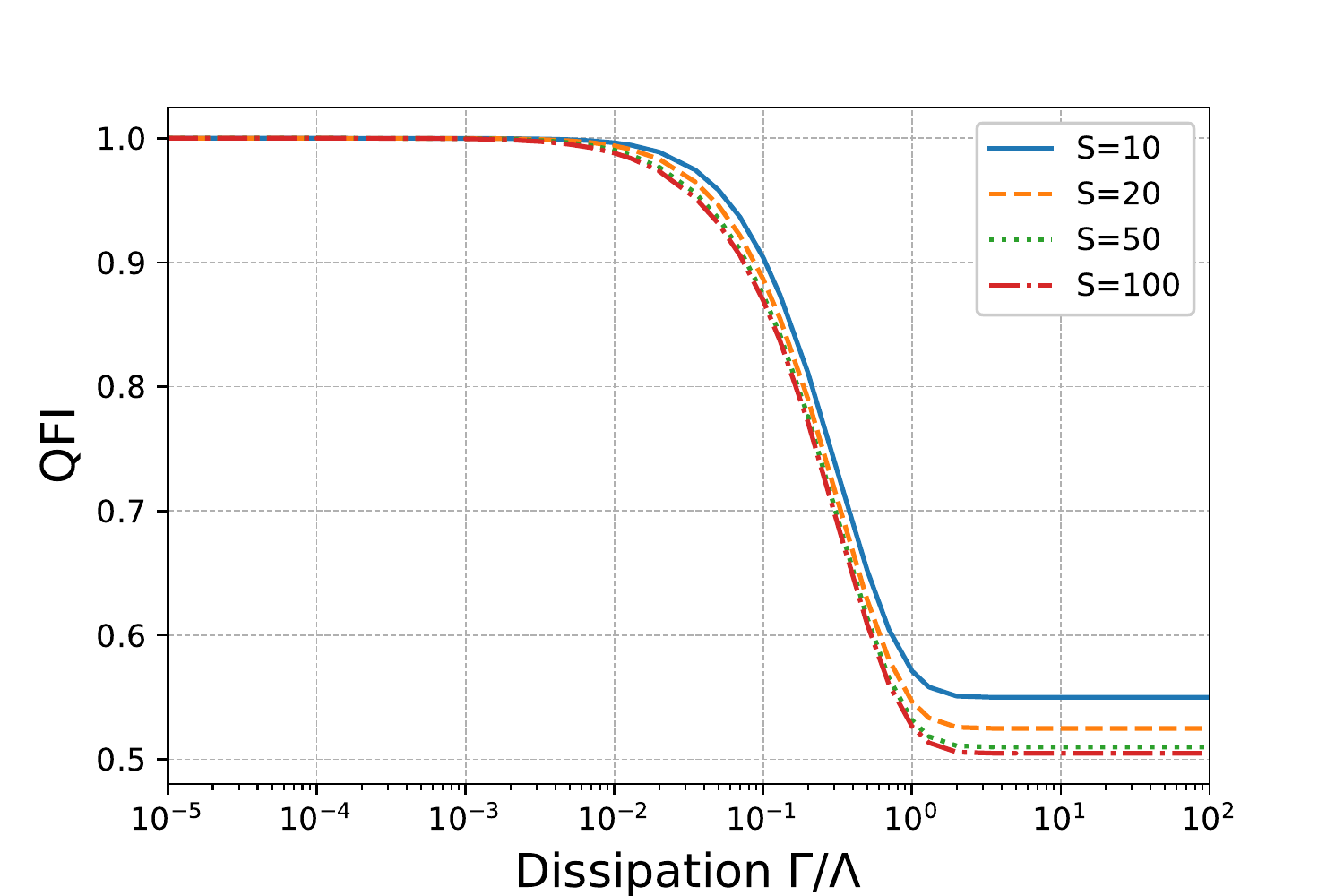}
	\caption{Fidelity (top) and QFI (bottom) as a function of the strength of dissipation $\Gamma /\Lambda$ at the time $t=\pi /(2\Lambda )$ given that no jump has occurred. The QFI is plotted relative to the theoretical maximum, $4S^2$.}
	\label{fidqfi}
\end{figure}

Focusing still on the case in which no jump occurs, Fig.~\ref{fidqfi} shows the fidelity and QFI of the prepared quantum state as a function of the spin length $S$ and the dissipation rate $\Gamma$. 
Interestingly, the QFI is significantly more robust than the fidelity with respect to increasing spin length. 
For example, at $\Gamma /\Lambda =10^{-2}$ the fidelity drops from $\sim 1$ to $\sim 0.8$ with an increase of $S$ from 10 to 100, whereas the QFI remains within 1\% of the theoretical maximum $4S^2$.
The fidelity and QFI both level out at constant values once $\Gamma /\Lambda\gtrsim 3-5$ and this appears largely independent of the spin length $S$.

\begin{figure}[t]
\centering
	\includegraphics[width=\linewidth]{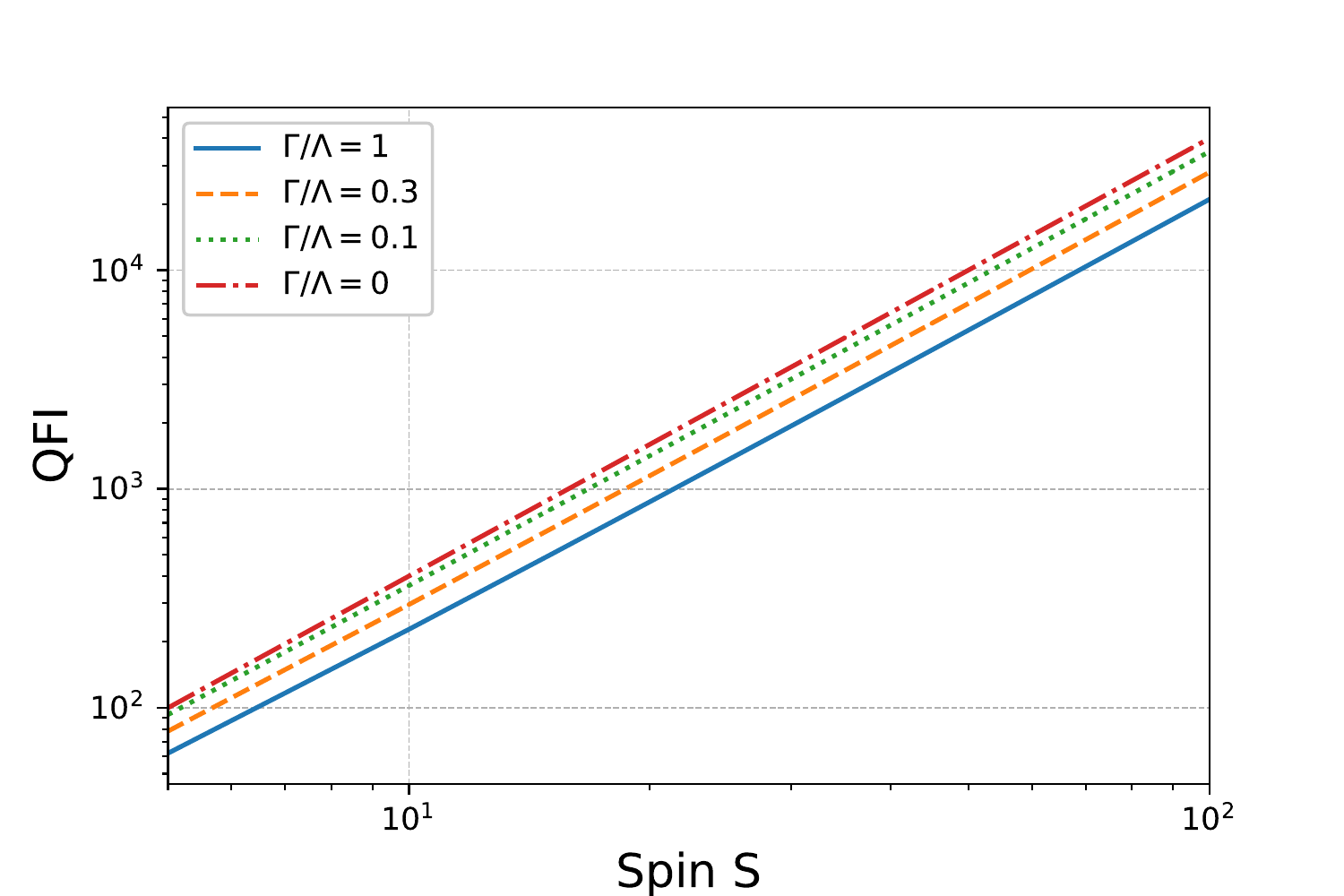}
	\caption{QFI of the quantum state at $t=\pi /(2\Lambda )$, given no jump has occurred, as a function of the spin length $S$ for a selection of  dissipation strengths $\Gamma /\Lambda$. The line for $\Gamma /\Lambda =0$ corresponds to the limit $4S^2$, while the line for $\Gamma /\Lambda =1$ corresponds approximately to the limit $2S^2+2S$.}
	\label{scaling}
\end{figure}

The optimal scaling of the QFI is quadratic: $4 S^2$. For increasing $\Gamma /\Lambda$ the QFI obviously decreases, and this is further illustrated in Fig.~\ref{scaling}, where we plot the QFI as a function of spin length $S$ for several values of $\Gamma /\Lambda$. 
We can in fact compute a lower bound for the QFI scaling in the absence of a jump, since we know that for sufficiently large $\Gamma /\Lambda$ (i.e., $\gtrsim 1$) the system ends up in the state $|S,0\rangle_x$, for which the QFI scales as $2S^2+2S$. This is still quadratic, and still clearly better than the scaling of standard quantum-limit states.

\subsection{Spontaneous Emission}
\label{singleparticledecoherence}
In the cavity QED framework, off-resonant excitation of atomic excited states by the Raman lasers leads to residual spontaneous emission that affects all atoms individually. In the atomic ground state manifold, the effects of this can be modelled by local dephasing ($\hat S^{(n)}_z$), excitation ($\hat S^{(n)}_+$), or deexcitation ($\hat S^{(n)}_-$), with the (no-jump) master equation modified to
\begin{equation}\label{SpE_ME}
\begin{split}
    \dot{\hat\rho}&=-i\left[\mathcal{\hat H},\hat\rho\right]     \\
    &+\frac{\gamma_{\rm eff}}{2}\sum_n^N\left(\mathcal{D}[\hat S^{(n)}_z]\hat\rho+\mathcal{D}[\hat S^{(n)}_+]\hat\rho+\mathcal{D}[\hat S^{(n)}_-]\hat\rho\right),
\end{split}
\end{equation}
where for simplicity we assume that the local effects all occur at the effective rate $\gamma_{\rm eff}$.

Again we can use the permutational invariance of the system, as described earlier and, also for simplicity, we consider just spin-1/2 atoms. With regards to the effects of spontaneous emission, we do not expect a significant difference between the spin-1/2 and spin-1 cases. 

In practice, we include both evolution with the collective and individual decay operators at the same time by performing in each time step a short evolution with the collective term followed by a short evolution with PIQS, 
\begin{equation}
    \hat\rho(t+\delta t)=e^{(\mathcal{L}_{\Lambda}+\mathcal{L}_{\gamma_{\rm eff}})\delta t}\left[e^{-\Gamma \hat S_x^2\delta t}\hat\rho(t)e^{-\Gamma \hat S_x^2\delta t}\right],
\end{equation}
where $\mathcal{L}_{\Lambda}$ corresponds to the Hermitian part of the Hamiltonian and $\mathcal{L}_{\gamma_{\rm eff}}$ is the Liouvillian operator corresponding to the second line of (\ref{SpE_ME}).
Note that due to the way in which PIQS separates the collective and local effects, it turns out that doing the whole evolution with the local effects and afterwards overlaying the evolution of the collective effects show the same final results.

\begin{figure}[H]
\centering
	\includegraphics[width=\linewidth]{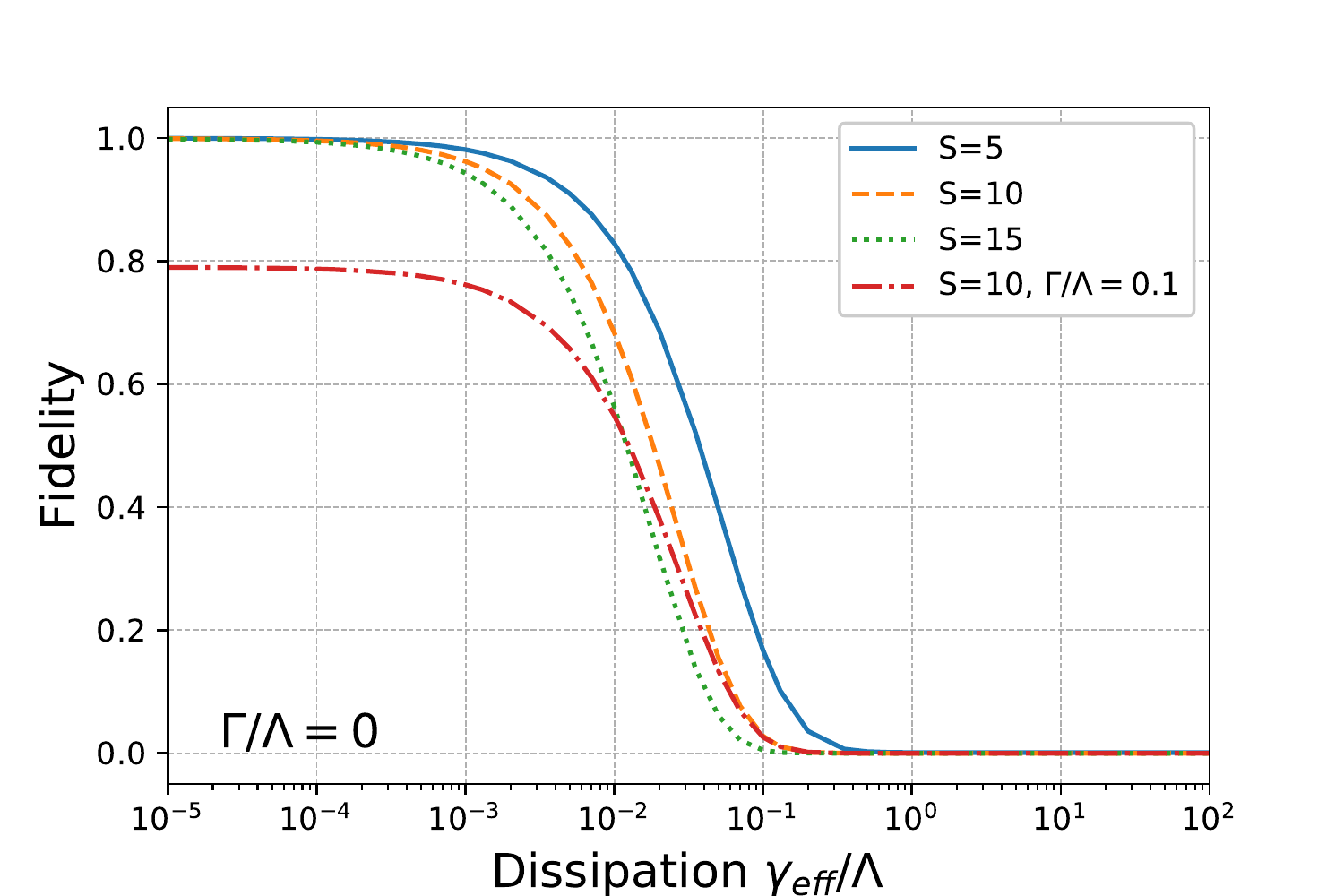}\\
	\includegraphics[width=\linewidth]{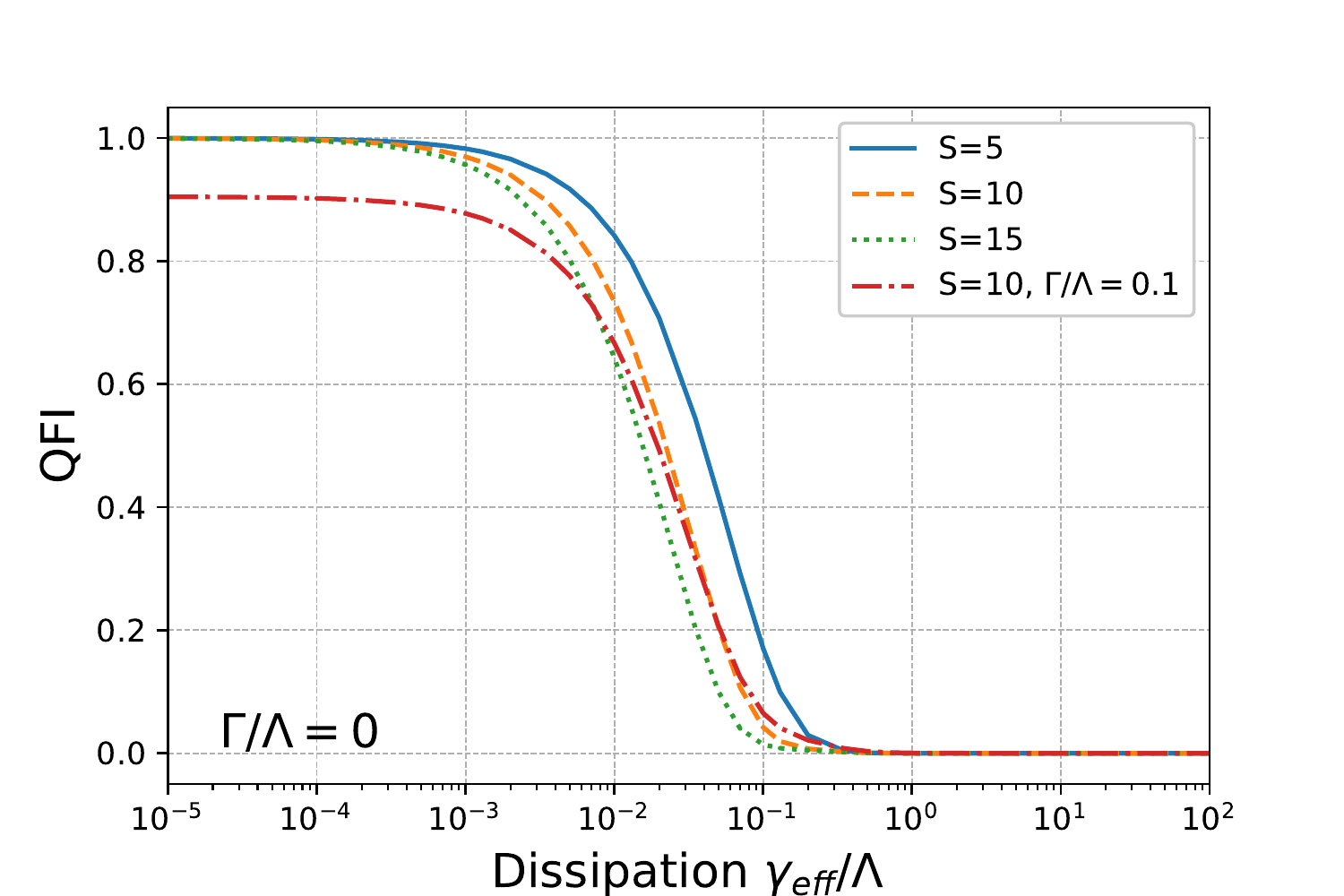}
	\caption{Fidelity (top) and QFI (bottom) as a function of the strength of dissipation $\gamma_{\rm eff}/\Lambda$ at the time $t=\pi /(2\Lambda )$, given that no jump (cavity photon emission) has occurred when $\Gamma >0$. The QFI is plotted relative to the theoretical maximum, $4S^2$.}
	\label{spem}
\end{figure}

Numerical results for the fidelity and QFI as a function of the dissipative rate $\gamma_{\rm eff}/\Lambda$ are shown in Fig.~\ref{spem}. Single-particle processes mix subspaces of different spin lengths, which increases the number of spin states available to the system and hence the size of the basis required for simulations. This means that our results are generally restricted to smaller spin lengths $S$. As can be seen in Fig.~\ref{spem}, the fidelity and QFI both show a greater sensitivity to $\gamma_{\rm eff}/\Lambda$ than to $\Gamma/\Lambda$, with the QFI now dropping off with $\gamma_{\rm eff}/\Lambda$ in a similar fashion to the fidelity. To maintain high values of the QFI clearly requires $\gamma_{\rm eff}/\Lambda <10^{-2}$. We note also that both the fidelity and the QFI go to zero once $\gamma_{\rm eff}/\Lambda \gtrsim 1$.


\subsection{Possible Experimental Parameters}
\label{exppar}
Considering the spin-1 realization of the Dicke model with ${}^{87}$Rb, the effective one-axis twisting parameter in the limit $\omega\gg\kappa$ is 
\begin{equation}
    \Lambda=\frac{g^2}{72\omega}\frac{\Omega^2}{\Delta^2} ,
\end{equation}
and $\Gamma=(\kappa/\omega)\Lambda$. Meanwhile, the effective rate of atomic spontaneous emission due to off-resonant excitation of the excited hyperfine level $5^2P_{1/2}$ is estimated to be
\begin{equation}
    \gamma_{\rm eff} \simeq \frac{\gamma}{12}\frac{\Omega^2}{\Delta^2} ,
\end{equation}
where $\gamma$ is the spontaneous emission line width of the $5^2P_{1/2}$ level. Hence, we have $\gamma_{\rm eff}/\Lambda\simeq 6\gamma\omega/g^2=(12/C)(\omega/\kappa)$, where $C=2g^2/(\kappa\gamma)$ is the single-atom cooperativity. Minimizing both $\Gamma/\Lambda$ and $\gamma_{\rm eff}/\Lambda$ therefore puts a very demanding requirement on the cooperativity $C$. For example, if we consider $\Gamma/\Lambda =\kappa/\omega=0.1$, then to achieve 
$\gamma_{\rm eff}/\Lambda =10^{-3}$ would require $C=120,000$. So, perhaps not unsurprisingly, it is very difficult to ensure Hamiltonian-dominated evolution in this cavity-QED-based scheme whilst also protecting the well-known fragility of the mesoscopic superposition states from the effects of spontaneous emission.

\subsection{Preparation of the Dicke state $|S,0\rangle_x$}

A more promising target for the cavity QED framework is in fact preparation of the Dicke state $|S,0\rangle_x$ (which also requires $S$ to be an integer), which still shows a quadratic scaling of the QFI with spin length ($2S^2+2S$). The generation of this specific Dicke state has been the subject of several investigations \cite{PhysRevA.99.023822,PhysRevLett.99.193602,PhysRevLett.112.155304,Zou6381,Luo620,PhysRevA.70.022106,RAGHAVAN2001149}.
As we saw earlier, with increasing spin length the probability to have no photon emission decreases with increasing $\Gamma$ or for large $t$, 
but there is still always a finite probability of having no cavity emissions, which in the absence of spontaneous emission heralds the preparation of the state $|S,0\rangle_x$. A simple formula for this probability can be obtained using Stirling's approximation ($S!\approx S^S e^S\sqrt{2\pi S}$) as \cite{PhysRevA.77.033810}
\begin{equation}
P_{m=0}=(D^S_{0,S})^2\approx\frac{1}{\sqrt{\pi S}}.
\end{equation}
So, the rate of decrease with $S$ is actually somewhat slow; e.g., for $S=30$ the probability is still 10\%. Note also that, as we shall see in the next section, if a photon emission does occur, then it triggers an ongoing sequence of photon emissions. So, in practice the distinction between preparation (no photon emission) and non-preparation (continual photon emission) of the state $|S,0\rangle_x$ in any particular run of the experiment should be clear. 

For this alternative target state, the preparation can be entirely dissipative in nature; that is, we can simply set $\omega=0$ (so $\Lambda=0$) and the rate at which the state is prepared is then determined by
\begin{equation}
\Gamma = \frac{g^2}{72\kappa}\frac{\Omega^2}{\Delta^2} .
\end{equation}
In particular, the $m\neq 0$ components of the initial state decay like $e^{-\Gamma m^2 t}$. Meanwhile, to avoid the effects of spontaneous emission we now only require $\gamma_{\rm eff}/\Gamma =12/C\ll 1$. Microcavity experiments with ${}^{87}$Rb atoms can already achieve $C\gtrsim 300$ \cite{PhysRevLett.104.203602} corresponding to $\gamma_{\rm eff}/\Gamma \lesssim 0.04$, while nanocavity experiments show promise of much larger cooperativities \cite{thompson2013coupling}.


\subsection{Jump trajectory: Entangled-State Cycles}
\label{jumps}
Let us assume that the cavity-mediated dynamics dominates over effects associated with spontaneous emission. 
By realizing that
\begin{equation}
    D^S_{-m,S}=D^S_{m,S} ,
\end{equation}
we can rewrite the wave function (\ref{wavefunction}) as
\begin{equation}
\begin{split}
    |\psi(t)\rangle&=\frac{\sum_m D^S_{m,S}e^{i\Lambda m^2 t}e^{-\Gamma m^2t}|S,m\rangle_x}{\sqrt{\sum_m e^{-2\Gamma m^2t}(D^S_{m,S})^2}}\\
    &=\frac{D^S_{0,S}|S,0\rangle_x+\sum_{m=1}^S D^S_{m,S}e^{-\Gamma m^2t}|\chi_S^+(m)\rangle}{\sqrt{\sum_m e^{-2\Gamma m^2t}(D^S_{m,S})^2}},
\end{split}    
\end{equation}
where $|\chi_S^\pm (m)\rangle$
are the ``kitten'' states
\begin{equation}
    |\chi_S^\pm(m)\rangle=\frac{e^{i\Lambda m^2 t}}{\sqrt{2}}\left(|S,m\rangle_x\pm|S,-m\rangle_x\right).
\end{equation}
Once a jump happens, the jump operator, being proportional to $\hat S_x$, makes it so that the $m=0$ component of the initial state vanishes and the $|\chi_S^\pm(m)\rangle$ states change their relative phase by $\pi$:
\begin{equation}
\label{jumpoperator}
    \hat S_x|\chi_S^\pm(m)\rangle=m|\chi_S^\mp(m)\rangle.
\end{equation}
The system then, after some time, settles probabilistically into one of $S$ possible cycles of jumps between the pairs of states $|\chi_S^\pm(m)\rangle$ (with an overall probability of $2(D^S_{m,S})^2$ for a given $m$) and the jumps between these two states continue {\em indefinitely} \cite{PhysRevA.77.033810}. 

\begin{figure}[H]
\centering
	\includegraphics[width=\linewidth]{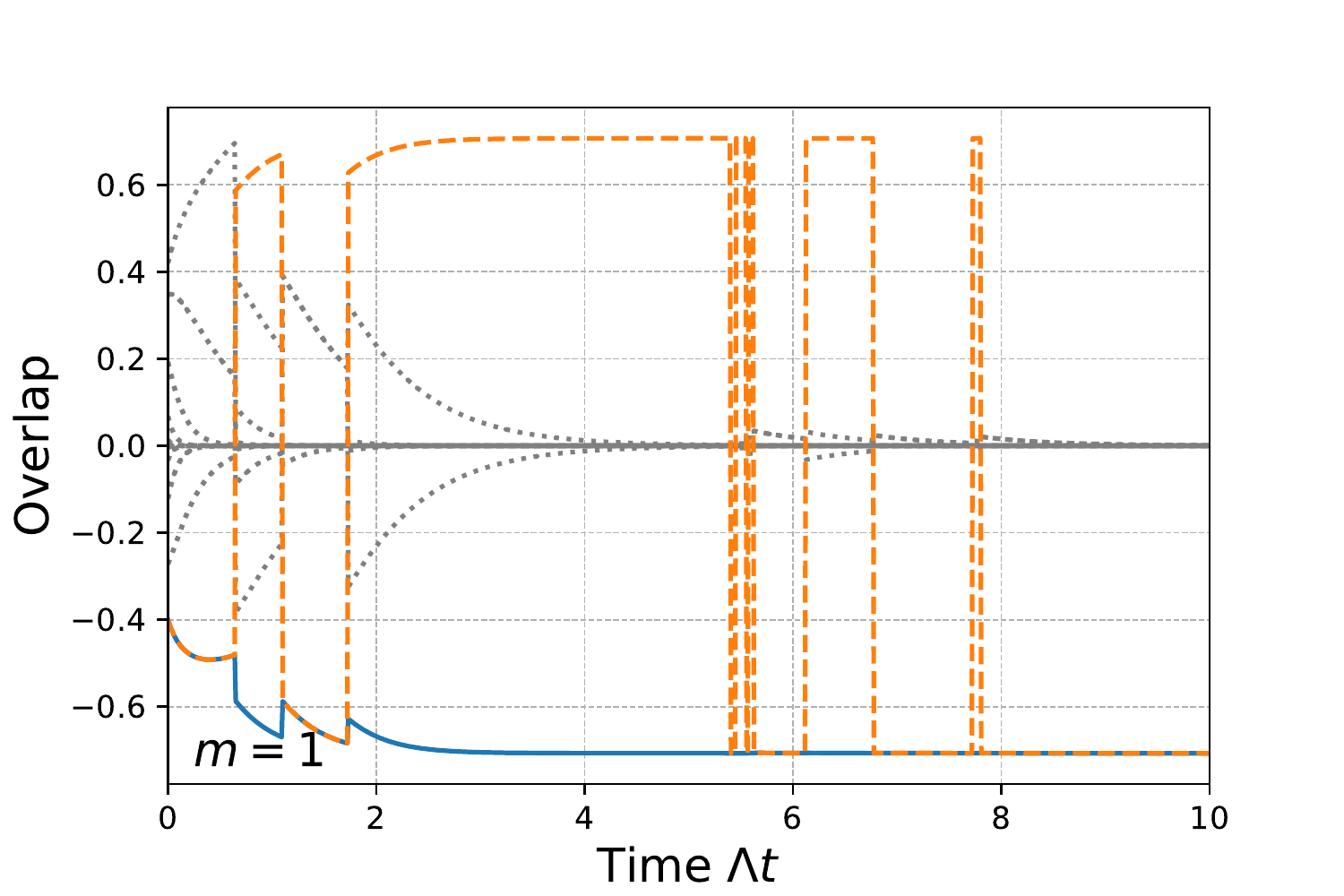}\\
	\includegraphics[width=\linewidth]{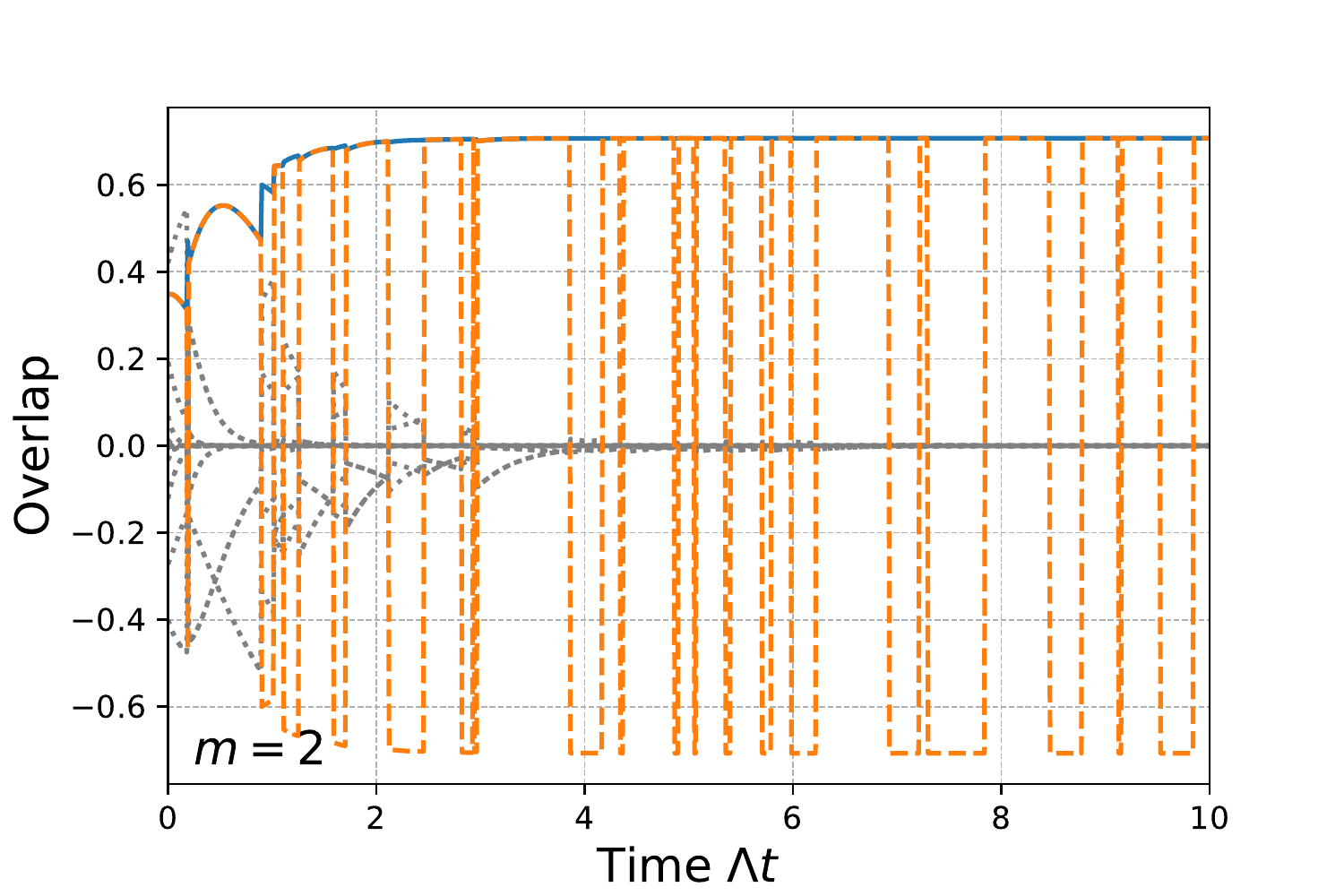}
	\caption{Monte Carlo simulations of the overlap of the system state with the eigenstates of $\hat S_x$. We plot $\langle\psi(t)|S,m\rangle_x$ (solid blue) and $\langle\psi(t)|S,-m\rangle_x$ (dashed orange) for the $m$-value that corresponds to the final, surviving entangled-state cycle, and the remaining overlaps (dotted grey) as a function of time for single trajectories of spin $S=10$, with $\Gamma /\Lambda=0.5$. The two trajectories show a cycle with $m=1$ (top) and $m=2$ (bottom). Note that we have removed the phase factor $e^{i\Lambda m^2 t}$ from all of the overlaps.}
	\label{overlapcycle}
\end{figure}

We can determine in which cycle the system has settled by observing the rate at which photons are emitted, as this rate depends quadratically on $m$, i.e.,
\begin{equation}
    \langle\chi_S^\pm(m)|\hat S_x^2|\chi_S^\pm(m)\rangle=m^2 .
\end{equation}
In Fig.~\ref{overlapcycle} we show a couple of examples of such jump trajectories. In particular, we plot the overlaps of the system state $|\psi(t)\rangle$ with the various eigenstates of $\hat S_x$. 
We see clearly how jumps lead to a relative sign change between the components $|S,\pm m\rangle_x$, and also how the frequency of jumps increases with increasing $m$.

The particular cycle in which the system ends up is random, but is in general strongly influenced by the time at which the first jump happens. An early jump may increase the relative amplitude of a higher-$m$ cycle above that of some lower ones, and therefore steer the system towards a higher-$m$ cycle, while the longer the first jump takes to happen, the more likely it is for  lower-$m$ cycles to be established, since only these remain with significant amplitudes. Note that the highest-$m$ cycles rarely appear, since their amplitudes are so suppressed that even an immediate jump would not make them significant. 
In fact, for $j$ immediate jumps, the relative amplitudes of the states $|S,m\rangle_x$ become
\begin{equation}
\begin{split}
P_{m}=(m^jD^S_{m,S})^2\approx\frac{m^{2j}}{\sqrt{\pi S}}e^{-\frac{m^2}{S}} ,
\end{split}
\end{equation}
which makes little difference for the largest $m$.

Once one $m$-cycle starts to dominate it is unlikely for the system to evolve to another cycle, as the dominant cycle determines the probability of jumps and hence the frequency (but this is not impossible, as can be seen in Fig.~\ref{overlapcycle} (bottom), where the system transitions from being a higher-$m$ cycle to the $m=2$-cycle after a small period of less frequent jumps). One essentially sees a positive feedback loop relationship between the jump frequency and the population of a specific entangled-state cycle.

Finally, we note that all of the results that we have presented assume an integer value of the total spin. If the individual atoms are taken to have spin-1/2, then one obviously requires an even number of atoms. For the typical numbers of atoms that we consider here, this could be achieved reliably with, e.g., an even-numbered array of optical tweezers, each tweezer containing one atom (similarly, in a trapped-ion configuration the number of ions can be precisely determined). However, the spin-1 realization of the Dicke model also offers a simple solution to this issue, as the total spin necessarily takes an integer value.


\section{Conclusion}
\label{concl}
 

To summarize, we have proposed a scheme for generating spin cat states using an engineered Dicke model, which represents a potential solution to the problem of generating entangled states in large ensembles of atoms. That one-axis twisting has the potential to create cat states was known before, but the novel engineering of it from a Dicke model creates a distinct model, where the non-jump part of the open-system (dissipative) evolution also takes the form of one-axis twisting.

The scheme can in principle be implemented in two different physical systems: trapped ions and cavity QED. We examined parameter regimes from state-of-the-art cavity and trapped-ion experiments. Of these two, the trapped ions appear to be more promising because of the vanishing collective noise from heating and the manageable single-particle dephasing, whereas the cooperativities required for a cavity QED implementation are very challenging, but hopefully still lie in the near future. However, the cavity QED setup does also have the benefit of offering the option to generate alternative but still interesting states, such as Schr\"odinger ``kitten'' states, and the Dicke $|S,0\rangle_x$. These states are also clearly identified via properties the cavity output field.


The usefulness and potential manipulation of the entangled-state cycles remains somewhat speculative. Future directions could be to look more closely at them, in which case one would also have to consider how single-particle decoherence processes affect the systems. Since single-particle operators produce jumps between different Dicke subspaces, the entangled-state cycles could evolve to superpositions made up of Dicke states from different subspaces. Other than that, the decay and pumping of individual spins would shift the magnetic quantum number of the two components of the kitten state by the same amount, thus creating superpositions of different $m$ which decay at different speeds, or even potentially removing the components with the maximal/minimal magnetic quantum number $m=\pm S$.


\begin{acknowledgments} 
This work makes use of the Quantum Toolbox in Python (QuTiP) \cite{JOHANSSON20121760,JOHANSSON20131234}.
\end{acknowledgments}

\appendix

\section{Adiabatic Elimination of the bosonic mode}
We move into the interaction picture via the transformation
\begin{equation}
    \hat\rho'(t)=e^{i(\omega\hat a^\dagger \hat a+\omega_0\hat S_z) t}\hat\rho(t)e^{-i(\omega\hat a^\dagger \hat a+\omega_0\hat S_z)t} ,
\end{equation}
which yields the interaction picture Hamiltonian
\begin{equation}
    \hat H_I(t)=\hat a e^{-i\omega t}\hat X(t)+\hat a^\dagger e^{i\omega t}\hat X^\dagger(t),
\end{equation}
where
\begin{equation}
    \hat X(t)=\frac{\lambda_-}{\sqrt{2S}}\hat S_-e^{-i\omega_0t}+\frac{\lambda_+}{\sqrt{2S}}\hat S_+e^{i\omega_0t}.
\end{equation}
Under the assumption that $\sqrt{\omega^2+\kappa^2}\gg\omega_0,\lambda_\pm$, we expand to second order in the interaction Hamiltonian and trace out the bosonic environment $E$ as follows,
\begin{widetext}
\begin{equation}
\begin{split}
    \dot{\hat\rho}_S'(t) &=-\int_0^tdt'\text{Tr}_E\left(\left[ \hat H_I(t),e^{\mathcal{L}_E(t-t')}\left[\hat H_I(t'),\hat\rho_S'\otimes\hat\rho_E'(t')\right]\right]\right)\\
    &=\int_0^tdt'\hat X^\dagger(t)\hat X(t')\hat\rho_S'(t')\text{Tr}_E\bigg[\hat ae^{\mathcal{L}_E(t-t')}\big(\hat a^\dagger\hat\rho_E'(t')\big)\bigg]e^{-i\omega(t-t')}\\
    &-\int_0^tdt'\hat X^\dagger(t)\hat\rho_S'(t')\hat X(t')\text{Tr}_E\bigg[\hat ae^{\mathcal{L}_E(t-t')}\big(\hat\rho_E'(t')\hat a^\dagger\big)\bigg]e^{-i\omega(t-t')}\\
    &-\int_0^tdt'\hat X(t')\hat\rho_S'(t')\hat X^\dagger(t)\text{Tr}_E\bigg[\hat ae^{\mathcal{L}_E(t-t')}\big(\hat a^\dagger\hat\rho_E'(t')\big)\bigg]e^{-i\omega(t-t')}\\
    &+\int_0^tdt'\hat\rho_S'(t')\hat X(t')\hat X^\dagger(t)\text{Tr}_E\bigg[\hat ae^{\mathcal{L}_E(t-t')}\big(\hat\rho_E'(t')\hat a^\dagger\big)\bigg]e^{-i\omega(t-t')}\\
    &+\int_0^tdt'\hat X(t)\hat X^\dagger(t')\hat\rho_S'(t')\text{Tr}_E\bigg[\hat a^\dagger e^{\mathcal{L}_E(t-t')}\big(\hat a\hat\rho_E'(t')\big)\bigg]e^{i\omega(t-t')}\\
    &-\int_0^tdt'\hat X(t)\hat\rho_S'(t')\hat X^\dagger(t')\text{Tr}_E\bigg[\hat a^\dagger e^{\mathcal{L}_E(t-t')}\big(\hat\rho_E'(t')\hat a\big)\bigg]e^{i\omega(t-t')}\\
    &-\int_0^tdt'\hat X^\dagger(t')\hat\rho_S'(t')\hat X(t)\text{Tr}_E\bigg[\hat a^\dagger e^{\mathcal{L}_E(t-t')}\big(\hat a\hat\rho_E'(t')\big)\bigg]e^{i\omega(t-t')}\\
    &+\int_0^tdt'\hat\rho_S'(t')\hat X^\dagger(t')\hat X(t)\text{Tr}_E\bigg[\hat a^\dagger e^{\mathcal{L}_E(t-t')}\big(\hat\rho_E'(t')\hat a\big)\bigg]e^{i\omega(t-t')} ,
\end{split}
\end{equation}
\end{widetext}
where 
\begin{equation}
\mathcal{L}_E\rho = \kappa(\bar{\mathfrak{n}}+1) \mathcal{D}[\hat a]\hat\rho+\kappa\bar{\mathfrak{n}} \mathcal{D}[\hat a^\dagger]\hat\rho ,
\end{equation}
and we have eliminated terms proportional to $\hat a\hat a$ and $\hat a^\dagger\hat a^\dagger$, because their expectation value for a thermal state is zero. Next, because of the separation of timescales between the system and the bosonic environment, we can set $\{ \hat X(t'),\hat X^\dag (t'),\hat\rho_S'(t')\}\rightarrow \{ \hat X(t),\hat X^\dag (t),\hat\rho_S'(t)\}$ inside the integrals. 

Given that the environment $E$ is in a thermal state, we evaluate
\begin{equation}
    \begin{split}
        \text{Tr}_E\bigg[\hat ae^{\mathcal{L}_E(t-t')}\big(\hat a^\dagger\hat\rho_E'(t')\big)\bigg]&=(\bar{\mathfrak{n}}+1)e^{-\kappa(t-t')}\\
        \text{Tr}_E\bigg[\hat a e^{\mathcal{L}_E(t-t')}\big(\hat\rho_E'(t')\hat a^\dagger\big)\bigg]&=\bar{\mathfrak{n}} e^{-\kappa(t-t')}\\
        \text{Tr}_E\bigg[\hat a^\dagger e^{\mathcal{L}_E(t-t')}\big(\hat a\hat\rho_E'(t')\big)\bigg]&=\bar{\mathfrak{n}}e^{-\kappa(t-t')}\\
        \text{Tr}_E\bigg[\hat a^\dagger e^{\mathcal{L}_E(t-t')}\big(\hat\rho_E'(t')\hat a\big)\bigg]&=(\bar{\mathfrak{n}}+1) e^{-\kappa(t-t')} ,
    \end{split}
\end{equation}
and the resulting integrals can be reduced to
\begin{equation}
    \int_0^tdt'e^{-(\kappa\pm i\omega)(t-t')}\approx\frac{1}{\kappa\pm i\omega}.
\end{equation}
After rotating back to the Schrödinger picture this leaves us with
\begin{equation}
    \begin{split}
        \dot{\hat\rho}_S&=-i\left[\hat H,\hat\rho_S\right]\\
        &+(\bar{\mathfrak{n}}+1)\bigg(-\frac{1}{\kappa+i\omega}\hat X^\dagger\hat X\hat\rho_S+\frac{1}{\kappa+i\omega}\hat X\hat\rho_S\hat X^\dagger\\
        &+\frac{1}{\kappa-i\omega}\hat X\hat\rho_S\hat X^\dagger-\frac{1}{\kappa-i\omega}\hat\rho_S\hat X^\dagger\hat X\bigg)\\
        &+\bar{\mathfrak{n}}\bigg(-\frac{1}{\kappa-i\omega}\hat X\hat X^\dagger\hat\rho_S+\frac{1}{\kappa+i\omega}\hat X^\dagger\hat\rho_S\hat X\\
        &+\frac{1}{\kappa-i\omega}\hat X^\dagger\hat\rho_S\hat X-\frac{1}{\kappa+i\omega}\hat\rho_S\hat X\hat X^\dagger\bigg)\\
        &=-i\left[\hat H,\hat\rho_S\right]\\
        &+\frac{\kappa(\bar{\mathfrak{n}}+1)}{\omega^2+\kappa^2}\left(2\hat X\hat\rho_S\hat X^\dagger-\hat X^\dagger\hat X\hat\rho_S-\hat\rho_S\hat X^\dagger\hat X\right)\\
       &+\frac{\kappa\bar{\mathfrak{n}}}{\omega^2+\kappa^2}\left(2\hat X^\dagger\hat\rho_S\hat X-\hat X\hat X^\dagger\hat\rho_S-\hat\rho_S\hat X\hat X^\dagger\right),
    \end{split}
\end{equation}
where
\begin{equation}
    \hat H=\omega_0\hat S_z-\frac{\omega(\bar{\mathfrak{n}}+1)}{\omega^2+\kappa^2}\hat X^\dagger\hat X+\frac{\omega\bar{\mathfrak{n}}}{\omega^2+\kappa^2}\hat X\hat X^\dagger
\end{equation}
and
\begin{equation}
    \hat X=\frac{\lambda_-}{\sqrt{2S}}\hat S_-+\frac{\lambda_+}{\sqrt{2S}}\hat S_+.
\end{equation}

\section{Derivation of the quantum state at $t=\pi /(2\Lambda)$}

For integer $S$, the unnormalized wave function is given by
\begin{equation}
\begin{split}
&e^{-i\mathcal{\hat H}t}|\psi(0)\rangle=\sum_m D^S_{m,S}e^{i\frac{\pi}{2} m^2}e^{-\frac{\pi\Gamma}{2\Lambda} m^2}|S,m\rangle_x\\
&=\sum_{m\text{ even}} D^S_{m,S}e^{-\frac{\pi\Gamma}{2\Lambda} m^2}|S,m\rangle_x\\
&+i\sum_{m\text{ odd}} D^S_{m,S}e^{-\frac{\pi\Gamma}{2\Lambda} m^2}|S,m\rangle_x\\
&=\sum_{m\text{ even}} \frac{1}{2}\left(D^S_{m,S}+(-1)^SD^S_{m,-S}\right)e^{-\frac{\pi\Gamma}{2\Lambda} m^2}|S,m\rangle_x\\
&+i\sum_{m\text{ odd}} \frac{1}{2}\left(D^S_{m,S}-(-1)^SD^S_{m,-S}\right)e^{-\frac{\pi\Gamma}{2\Lambda} m^2}|S,m\rangle_x\\
&=\sum_{m} \frac{1}{2}\left(D^S_{m,S}+(-1)^SD^S_{m,-S}\right)e^{-\frac{\pi\Gamma}{2\Lambda} m^2}|S,m\rangle_x\\
&+i\sum_{m} \frac{1}{2}\left(D^S_{m,S}-(-1)^SD^S_{m,-S}\right)e^{-\frac{\pi\Gamma}{2\Lambda} m^2}|S,m\rangle_x\\
&-\sum_{m\text{ odd}} \frac{1}{2}\underbrace{\left(D^S_{m,S}+(-1)^SD^S_{m,-S}\right)}_{=0}e^{-\frac{\pi\Gamma}{2\Lambda} m^2}|S,m\rangle_x\\
&-i\sum_{m\text{ even}} \frac{1}{2}\underbrace{\left(D^S_{m,S}-(-1)^SD^S_{m,-S}\right)}_{=0}e^{-\frac{\pi\Gamma}{2\Lambda} m^2}|S,m\rangle_x\\
&=\frac{1+i}{2}\sum_m D^S_{m,S}e^{-\frac{\pi\Gamma}{2\Lambda} m^2}|S,m\rangle_x\\
&~~~+(-1)^S\frac{1-i}{2}\sum_m D^S_{m,-S}e^{-\frac{\pi\Gamma}{2\Lambda} m^2}|S,m\rangle_x
\end{split}
\end{equation}
where in multiple places we have used
\begin{equation}
\label{Dmatrix}
D^S_{m,S}=(-1)^{S-m}D^S_{m,-S}.
\end{equation}

\clearpage

\bibliographystyle{apsrev4-1}
\bibliography{bib}

\end{document}